\documentclass[english,aps,prc,nofootinbib,superscriptaddress,twocolumn,letter]{revtex4}
\usepackage[latin9]{inputenc}
\usepackage{hyperref}
\hypersetup{
  colorlinks=true,        
  linkcolor=blue,         
  citecolor=cyan,         
}
\usepackage{breakurl}
\usepackage{graphicx}
\usepackage{epsf}
\usepackage{epsfig}
\usepackage{amssymb,amsmath}
\usepackage[usenames]{color}
\usepackage{amssymb}
\usepackage{times}
\usepackage{comment}

\newcommand\pt{p_\text{T}}

\allowdisplaybreaks



\begin{document}

\preprint{This line only printed with preprint option}

\title{Magnetization and Magnetic Field-Induced Correction: Implications for QGP Thermal Photon Production in Magnetohydrodynamic}

\author{Jing Jing}
\affiliation{Hubei Engineering University, Xiaogan, Hubei, 432000, China}

\author{Duan She}
\affiliation{Institute of Physics, Henan Academy of Sciences, Zhengzhou 450046, China}
\affiliation{Key Laboratory of Quark and Lepton Physics (MOE), Central China Normal University, Wuhan, Hubei, 430079, China}

\author{Ze-Fang Jiang}
\email{jiangzf@mails.ccnu.edu.cn}
\affiliation{Hubei Engineering University, Xiaogan, Hubei, 432000, China}

\affiliation{Key Laboratory of Quark and Lepton Physics (MOE), Central China Normal University, Wuhan, Hubei, 430079, China}

\begin{abstract}
We investigate thermal photon emission from magnetized quark-gluon plasma (QGP) within (1+1)-dimensional relativistic magnetohydrodynamics (MHD), systematically incorporating magnetic susceptibility $\chi_m$---encompassing both constant and lattice-QCD-derived temperature-dependent $\chi_m(T)$ parametrizations---and weak-field quantum corrections to quark distribution functions $f_{\rm EM}$. Employing the Pu-Bjorken MHD framework, we calculate photon production rates from Compton scattering, $q\bar{q}$ annihilation, bremsstrahlung, and annihilation with rescattering, and integrate these over the QGP spacetime evolution to obtain transverse momentum ($p_T$) spectra. Our results demonstrate that photon yields are predominantly governed by the initial magnetic field strength and its temporal decay profile, with $\chi_m$ exerting negligible influence in the explored parameter space. In contrast, the weak-field correction $f_{\rm EM}$ induces a distinct enhancement in thermal photon production at intermediate $p_T$. This work establishes a rigorous theoretical framework for quantifying electromagnetic observables in magnetized QGP and provides the foundation for future dissipative MHD studies incorporating spin-magnetization dynamics.

\end{abstract}


\maketitle

\section{Introduction}

Ultra-relativistic heavy-ion collisions at the Relativistic Heavy Ion Collider (RHIC) and the Large Hadron Collider (LHC) create a deconfined quark-gluon plasma (QGP) under extreme temperature and density conditions, providing a unique environment to study strongly interacting matter in the laboratory~\cite{STAR:2005gfr,ALICE:2008ngc}. In non-central collisions, the relativistic motion of charged spectator nucleons generates transient ultra-strong magnetic fields of magnitude $10^{18}$--$10^{20}$~G~\cite{Deng:2012pc,Li:2016tel,Gursoy:2014aka,Huang:2022qdn}. These fields are predicted to induce profound modifications to QGP dynamics through quantum anomalies---notably the chiral magnetic effect (CME) and chiral separation effect (CSE)---and through alterations of bulk transport properties~\cite{Kharzeev:2007jp,Fukushima:2008xe,Kharzeev:2015znc,Gursoy:2018yai,Jiang:2022uoe}. However, the experimental isolation of these magnetic-field-induced phenomena remains challenging due to substantial backgrounds from collective flow~\cite{STAR:2015wza,CMS:2017lrw,STAR:2021mii}.

Relativistic hydrodynamics provides the principal theoretical framework for modeling QGP evolution, successfully describing key observables including harmonic flow and global spin polarization~\cite{Heinz:2013th,Gale:2013da,Becattini:2017gcx,Jiang:2021ajc,Zhao:2022ayk}. To capture the interplay between electromagnetic fields and the QGP, the (3+1)-dimensional relativistic magnetohydrodynamic (MHD) equations must be solved self-consistently~\cite{Inghirami:2016iru,Nakamura:2022idq,Mayer:2024kkv}. While lattice QCD calculations constrain the temperature-dependent electrical conductivity and magnetic susceptibility $\chi_m$ of the QGP~\cite{Ding:2010ga,Bali:2013owa,Ding:2016hua}, the coupled spatiotemporal evolution of initial magnetic fields and the medium remains incompletely characterized~\cite{Pang:2016yuh,Jiang:2024bez,Huang:2024aob}.

Thermal photons serve as penetrating probes of the QGP, emitted throughout the fireball evolution and escaping without significant final-state interaction, thereby encoding direct information about the thermodynamic history and electromagnetic response~\cite{Gale:2003iz,Bhatt:2010cy,vanHees:2011vb,Shen:2013vja,Paquet:2015lta,Wang:2020dsr,Dwibedi:2025xho,Xiong:2025koa}. The dominant production mechanisms---Compton scattering and $q\bar{q}$ annihilation (C+A), bremsstrahlung (Bre), and $q\bar{q}$ annihilation with additional scattering (A+S)---exhibit characteristic transverse-momentum ($p_T$) dependence that carries signatures of parton dynamics in magnetized plasmas~\cite{Traxler:1995kx,Steffen:2001pv}. While relativistic MHD has been applied to magnetized QGP evolution in ideal Bjorken flow~\cite{Roy:2015kma,Pu:2016ayh}, longitudinal expansion~\cite{She:2019wdt}, and rotating systems~\cite{Shokri:2018qcu}, critical gaps persist: existing treatments rarely incorporate both the magnetic susceptibility $\chi_m$~\cite{Bali:2014kia,Bali:2020bcn}---which quantifies the medium's magnetization response---and weak-field quantum corrections to quark distribution functions ($f_{\rm EM}$)~\cite{Sun:2023pil,Sun:2023rhh,Sun:2024isb} within a unified framework for photon production. Consequently, the impact of $\chi_m$ on electromagnetic observables remains unexplored, leaving unresolved how the magnetic response modulates thermal photon yields.

To address these limitations, we present three principal advancements in this work: (i) the implementation of $\chi_m$-dependent ideal MHD using both constant values and lattice-QCD-constrained temperature-dependent parametrizations [$\chi_{m}^{\rm 2014}(T)$~\cite{Bali:2014kia} and $\chi_{m}^{\rm 2020}(T)$~\cite{Bali:2020bcn}]; (ii) the incorporation of $f_{\rm EM}$ corrections accounting for magnetic-field-modified quark distributions~\cite{Sun:2023pil}; and (iii) the systematic quantification of photon yields as functions of $\chi_m$ parametrizations, magnetic field decay exponents ($a$), and initial field strengths ($\sigma$). Building upon the magnetized Bjorken flow framework~\cite{Roy:2015kma,Pu:2016ayh,Jiang:2024mts}, we derive analytical solutions for the $\chi_m$-dependent temperature evolution and integrate thermal photon production rates over the full spacetime history to obtain $p_T$ spectra. This approach enables the disentanglement of primary regulators ($a$, $\sigma$) from secondary quantum and medium effects ($\chi_m$, $f_{\rm EM}$) across different $p_T$ regimes and evolution stages.

Our work extends previous investigations of photon production in viscous~\cite{Steffen:2001pv,Bhatt:2010cy}, accelerated~\cite{Jiang:2020big,She:2019wdt,Kasza:2025wot,HaddadiMoghaddam:2020ihi}, and magnetohydrodynamic~\cite{Roy:2015kma,Pu:2016ayh,Kushwah:2025jsb,Xiong:2025koa} fluids into a fully consistent MHD framework that incorporates arbitrary magnetic field strengths, establishing a benchmark for constraining QGP magnetic properties. Photon emission in magnetized QGP has previously been studied at weak coupling, notably via the synchrotron radiation formalism~\cite{Tuchin:2012mf,Tuchin:2014pka} (see also Refs.~\cite{Zakharov:2016mmc,Hattori:2016cnt,Wang:2024gnh}), and at strong coupling via holographic methods~\cite{Yee:2013qma,Muller:2013ila, Wu:2013qja}. These studies, which employ complementary theoretical approaches, provide useful context for the weak-field perturbative calculation developed here.
The results provide testable predictions for heavy-ion experiments, guiding the extraction of electromagnetic response signatures from data.

This paper is organized as follows. Section~\ref{sec:2} presents the MHD framework, $\chi_m$ prescriptions, $f_{\rm EM}$ corrections, and the photon production rate formalism. Section~\ref{section-4} reports the thermal photon spectra and analyzes the parameter dependence. Section~\ref{section-5} summarizes our conclusions and outlines future directions. We employ the Minkowski metric $g^{\mu\nu}={\rm diag}(1,-1,-1,-1)$, with fluid four-velocity $u^\mu$ normalized as $u^\mu u_\mu=1$ and the spatial projection operator $\Delta^{\mu\nu}=g^{\mu\nu}-u^\mu u^\nu$.

\section{Formalism}
\label{sec:2}

\subsection{Magnetohydrodynamic Framework}
\label{sec:2-A}

We consider the (1+1)-dimensional ideal magnetohydrodynamic (MHD) framework---Pu-Bjorken flow~\cite{Pu:2016ayh}---to investigate magnetic susceptibility effects on quark-gluon plasma (QGP) expansion dynamics in heavy-ion collisions. The total energy-momentum tensor for a relativistic magnetized fluid assumes the form
\begin{equation}
\begin{aligned}
T^{\mu\nu} =& (\varepsilon + p - MB + B^{2})u^{\mu}u^{\nu} - \left(p - MB + \frac{1}{2}B^{2}\right)g^{\mu\nu} \\
&+ (MB - B^{2})b^{\mu}b^{\nu},
\label{tmunu_total}
\end{aligned}
\end{equation}
where $\varepsilon$ and $p$ denote the fluid energy density and thermodynamic pressure, respectively, and $u^{\mu}$ is the fluid four-velocity. The magnetic field four-vector $B^\mu$ and magnetization four-vector $M^\mu$ are defined, respectively, as
\begin{equation}
B^{\mu} = \frac{1}{2}\epsilon^{\mu\nu\alpha\beta}u_{\nu}F_{\alpha\beta}, \quad M^{\mu} = \chi_{m}B^{\mu},
\label{eq:B_M_def}
\end{equation}
with $F^{\mu\nu} = \partial^\mu A^\nu - \partial^\nu A^\mu$ denoting the Faraday tensor and $\epsilon^{\mu\nu\alpha\beta}$ the Levi-Civita tensor ($\epsilon^{0123} = +1$). These definitions imply the orthogonality condition $u^\mu B_\mu = 0$, the Lorentz-invariant magnetic field strength $B = \sqrt{-B^\mu B_\mu}$, the magnetization magnitude $M = \chi_m B$, and the spacelike unit vector $b^{\mu} = B^{\mu}/B$ satisfying $b^\mu b_{\mu} = -1$.

In the ideal MHD limit, the comoving electric field vanishes ($E^\mu = 0$), ensuring a finite charge current $j^\mu = \sigma_{\rm el} E^\mu$~\cite{Roy:2015kma}. The magnetic field evolution is governed by the covariant frozen-flux theorem~\cite{Roy:2015kma}:
\begin{equation}
\partial_\nu (B^\mu u^\nu - B^\nu u^\mu) = 0,
\label{eq:frozen_flux}
\end{equation}
which implies that magnetic field lines are advected with the fluid and cannot diffuse through the conducting medium. The system is closed by energy-momentum conservation:
\begin{equation}
\partial_{\mu}T^{\mu\nu} = 0.
\label{eq:EM_cons}
\end{equation}

For the high-temperature deconfined QGP, we adopt the conformal equation of state $p = c_s^2 \varepsilon = \varepsilon/\kappa$, where $\kappa = 1/c_s^2$ denotes the equation-of-state parameter. For the ultrarelativistic QGP, $\kappa = 3$~\cite{Heinz:2013th}, consistent with a deconfined phase dominated by massless partonic degrees of freedom. We note that lattice QCD calculations suggest a temperature-dependent $c_s(T)$ for more realistic descriptions~\cite{Huovinen:2009yb}. Crucially, and in contrast to earlier treatments that neglected magnetization~\cite{Roy:2015kma}, we incorporate QGP magnetization~\cite{Pu:2016ayh} to account for magnetic-field-induced modifications to the fluid thermodynamic state.

We consider longitudinally boost-invariant flow, the standard framework for describing midrapidity dynamics in high-energy heavy-ion collisions where the fireball expands primarily along the beam ($z$) direction. In Milne coordinates, defined by $t = \tau \cosh\eta_s$ and $z = \tau \sinh\eta_s$ with proper time $\tau = \sqrt{t^2 - z^2}$ and spacetime rapidity $\eta_s = \frac{1}{2}\ln[(t+z)/(t-z)]$, the fluid four-velocity becomes
\begin{equation}
u^{\mu} = (\cosh\eta_s, 0, 0, \sinh\eta_s) = \gamma\left(1, 0, 0, \frac{z}{t}\right),
\label{eq:u_mu}
\end{equation}
where $\gamma = \cosh\eta_s$ denotes the Lorentz factor. The fundamental differential operators in this coordinate system are
\begin{equation}
u^{\mu} \partial_{\mu} = \frac{\partial}{\partial\tau}, \quad \partial_{\mu}u^{\mu} = \frac{1}{\tau}.
\label{eq:diff_ops}
\end{equation}

Projecting Eq.~(\ref{eq:EM_cons}) onto $u_\mu$ (Landau-Lifshitz frame) yields the energy conservation equation. Utilizing the orthogonality condition $u_{\mu}b^{\mu} = 0$ and the Maxwell equation for ideal MHD~\cite{Pu:2016ayh}:
\begin{equation}
\frac{1}{2}(u^{\alpha}\partial_{\alpha})B^{2} + B^{2}\partial_{\alpha}u^{\alpha} + B^{2}b^{\mu}b^{\nu}\partial_{\nu}u_{\mu} = 0,
\label{eq:Maxwell_B}
\end{equation}
we obtain the energy conservation equation:
\begin{equation}
\frac{\partial \varepsilon }{\partial\tau} + \frac{\varepsilon + p - MB + B^{2}}{\tau} + \frac{1}{2} \frac{\partial B^2}{\partial\tau} = 0.
\label{eq:energy_cons}
\end{equation}
This equation explicitly incorporates magneto-fluid energy exchange through the $\partial B^2/\partial\tau$ term and magnetization effects via the $MB$ contribution. Further details regarding the derivation can be found in Ref.~\cite{Pu:2016ayh}.

Projecting Eq.~(\ref{eq:EM_cons}) orthogonal to $u^\mu$ via the spatial projector $\Delta_{\mu\nu} = g_{\mu\nu} - u_\mu u_\nu$ yields the momentum conservation equation:
\begin{equation}
\begin{aligned}
&(\varepsilon + p - MB + B^{2})u^{\mu}\partial_{\mu}u_{\alpha} - \Delta_{\mu\alpha}\partial^{\nu}\left(p - MB + \frac{1}{2}B^{2}\right) \\
&+ \Delta_{\mu\alpha}\partial_{\mu}\left[(MB - B^{2})b^{\mu}b^{\nu}\right] = 0.
\end{aligned}
\label{eq:mom_cons_full}
\end{equation}
For homogeneous transverse magnetic fields, where spatial gradients of $B^\mu$ vanish, the final term vanishes, reducing the momentum conservation equation to:
\begin{equation}
(\varepsilon + p - MB + B^{2})\partial_{\tau}u_{\alpha} - \Delta_{\nu\mu}\partial^{\nu}\left(p - MB + \frac{1}{2}B^{2}\right) = 0.
\label{eq:meq1}
\end{equation}
For the $\eta_s$ component, Eq.~(\ref{eq:meq1}) yields $\partial_{\eta_s}(p - MB + B^{2}/2) = 0$, confirming the Bjorken scaling property whereby thermodynamic variables depend exclusively on the proper time $\tau$. For the transverse directions ($\mu = x,y$), the spatial uniformity of pressure and magnetic field causes the derivative terms to vanish, which gives rise to geodesic motion with $\partial_\tau u_i = 0$:
\begin{equation}
\partial_{\tau}u_{i} - \frac{1}{\varepsilon + p - MB + B^{2}}\partial_{i}\left(p - MB + \frac{1}{2}B^{2}\right) = 0.
\label{eq:mom_transverse}
\end{equation}

In relativistic heavy-ion collisions, the magnetic field is transverse to the beam axis, generated by Lorentz-contracted spectator nucleons~\cite{Kharzeev:2007jp}. We orient the field along the $y$-direction, $\mathbf{B} = B\mathbf{e}_y$, homogeneous in the transverse ($x$-$y$) plane. In Milne coordinates, this transverse configuration is invariant under longitudinal boosts, and the frozen-flux theorem admits the power-law solution~\cite{Roy:2015kma,Pu:2016ayh,She:2019wdt,Biswas:2020rps}:
\begin{equation}
\mathbf{B}(\tau) = \mathbf{B}_0 \left(\frac{\tau_0}{\tau}\right)^a,
\label{eq:B_decay}
\end{equation}
where $a > 0$ parametrizes the decay rate, $\tau_0$ denotes the initial proper time, and $\mathbf{B}_0 = \mathbf{B}(\tau_0)$ represents the initial field strength. This form captures the advection of magnetic flux with the expanding fluid, analogous to the frozen-flux behavior in astrophysical plasmas~\cite{Baiotti:2016qnr}.

In the high-temperature, ideal-gas limit, the equation of state for the quark-gluon plasma reduces to the conformal form $p = c_{s}^{2}\varepsilon=a_1 T^4$~\cite{Muronga:2001zk}, where
\begin{equation}
a_1 = \left(16 + \frac{21}{2}N_f\right)\frac{\pi^2}{90}
\end{equation}
is determined by the number of quark flavors $N_f$ and gluon degrees of freedom~\cite{Muronga:2003ta}. The initial magnetic field strength is characterized by the dimensionless parameters~\cite{Roy:2015kma,Pu:2016ayh}
\begin{equation}
B_0^2 = \sigma_0 \varepsilon_0 = \sigma T_0^4,
\label{eq:b0_def}
\end{equation}
where $T_0 = T(\tau_0)$ denotes the initial temperature, $\varepsilon_0 = \varepsilon(\tau_0)$ the initial energy density, $\sigma_0 = B_0^2/\varepsilon_0$ the ratio of magnetic to fluid energy density, and $\sigma = B_0^2/T_0^4$ the field strength normalized to the temperature scale.

For clarity, we clarify the relation between the dimensionless parameter $\sigma$ and the physical magnetic field strength $|eB|/m_\pi^2$ commonly quoted in the heavy-ion literature. Using $T_0 = 0.31$~GeV and $m_\pi = 0.135$~GeV, one has $T_0^2 \approx 4.9\,m_\pi^2$. The magnetic field in natural units is related to $\sigma$ by $|eB_0| = e\sqrt{\sigma}\,T_0^2$, with $e = \sqrt{4\pi\alpha_{\rm EM}} \approx 0.303$. Therefore, $\sigma$ and $|eB_0|/m_\pi^2$ are connected by
\begin{equation}
\frac{|eB_0|}{m_\pi^2} \approx 1.48 \times \sqrt{\sigma},
\label{eq:sigma_to_eb}
\end{equation}

Thus, $\sigma = 0.01$, $0.1$, $1$, $10$ correspond to $|eB_0|/m_\pi^2 \approx 0.15$, $0.47$, $1.48$, $4.7$, respectively. At the hydrodynamic initial time $\tau_0 \approx 0.5$~fm/$c$, the vacuum-decay solution predicts $|eB|/m_\pi^2 \lesssim 10^{-2}$ at the fireball center~\cite{Yan:2021zjc,Stewart:2021mjz}, which corresponds to $\sigma \lesssim 5\times 10^{-5}$. However, if the QGP possesses a substantial electrical conductivity during the pre-equilibrium stage, the magnetic-field decay is significantly slowed~\cite{Huang:2022qdn}, and $\sigma$ at $\tau_0$ can reach $O(10^{-2})$--$O(10^{-1})$, i.e., $|eB_0|/m_\pi^2 \sim 0.1$--$0.5$. The values $\sigma = 1$--$30$ shown in several figures below span a deliberately broad parameter range---from the realistic weak-field regime ($\sigma \ll 1$) to the strong-field limit---in order to establish systematic trends and to illustrate the sensitivity of the framework. In all cases the weak-field condition $|eB| \ll T^2$ (equivalently $\sqrt{\sigma} \ll T_0^2 / m_\pi^2 \approx 4.9$) holds for $\sigma \ll 24$, ensuring the validity of the perturbative expansion used to derive $f_{\rm EM}$. We shall return to the implications of realistic $\sigma$ values in Sec.~\ref{section-4}.

\subsection{Temperature Evolution with Magnetic Susceptibility}
\label{sec:2-B}

Commencing from the energy conservation equation~(\ref{eq:energy_cons}), and incorporating the power-law magnetic field decay~(\ref{eq:B_decay}) together with the thermodynamic relations, we derive the evolution equation for temperature:
\begin{equation}
\frac{\partial T}{\partial \tau} + \frac{(\kappa+1)T}{4\kappa \tau} + (1 - a - \chi_{m})\frac{\sigma T_{0}^{4}\tau_{0}^{2a}}{4\kappa a_{1} T^{3} \tau^{2a+1}} = 0.
\label{eq:mhd-0}
\end{equation}
The first term represents the local rate of temperature variation, the second accounts for longitudinal expansion consistent with Bjorken scaling, and the third captures the magneto-fluid energy exchange, modulated by the magnetic field strength $\sigma$, decay exponent $a$, and magnetic susceptibility $\chi_m$.

\subsubsection{Constant Magnetic Susceptibility ($\chi_m = {\rm const.}$)}

Analytical integration of Eq.~(\ref{eq:mhd-0}) yields the exact solution for the temperature evolution:
\begin{equation}
\begin{aligned}
T = T_{0}\Bigg[\left(\frac{\tau_{0}}{\tau}\right)^{\frac{\kappa+1}{\kappa}} 
&+ \frac{\sigma(1 - a - \chi_m)}{a_{1}(\kappa(2a - 1) - 1)}\Bigg(\left(\frac{\tau_{0}}{\tau}\right)^{2a} \\
&\quad - \left(\frac{\tau_{0}}{\tau}\right)^{\frac{\kappa+1}{\kappa}}\Bigg)\Bigg]^{\frac{1}{4}}.
\label{T_mhd_1}
\end{aligned}
\end{equation}
This solution comprises two distinct contributions: the standard Bjorken-like cooling term scaling as $(\tau_0/\tau)^{(\kappa+1)/\kappa}$, and a magnetization correction that depends on $\chi_m$, $\sigma$, and $a$. We examine two physically relevant limits that connect to established results in the literature~\cite{Pu:2016ayh,Roy:2015kma}:

\emph{Case A: Ideal MHD limit ($a = 1$).} For infinite electrical conductivity, the magnetic field is frozen into the fluid and strictly advected with the flow. Substituting $a=1$ and $\kappa = 3$ into Eq.~(\ref{T_mhd_1}), we obtain:
\begin{equation}
T = T_{0}\left[\left(\frac{\tau_{0}}{\tau}\right)^{\frac{4}{3}} 
- \frac{\sigma\chi_m}{2a_{1}}\left(\left(\frac{\tau_{0}}{\tau}\right)^{2} - \left(\frac{\tau_{0}}{\tau}\right)^{\frac{4}{3}}\right)\right]^{\frac{1}{4}}.
\label{T_mhd_2}
\end{equation}
Here, positive $\chi_m$ (paramagnetic QGP) retards the temperature decay, as the fluid absorbs energy from the decaying magnetic field, whereas negative $\chi_m$ (diamagnetic hadronic phase) accelerates cooling, as the fluid expends energy to expel the magnetic flux. This result recovers the established Pu-Bjorken flow solution~\cite{Pu:2016ayh}.

\emph{Case B: Critical decay parameter ($a \to 2/3$).} When the denominator $\kappa(2a-1)-1$ vanishes for $\kappa=3$ (i.e., $a=2/3$), we apply L'Hopital's rule to resolve the indeterminate form in Eq.~(\ref{T_mhd_1}), yielding a logarithmic correction:
\begin{equation}
T = T_{0}\left(\frac{\tau_{0}}{\tau}\right)^{\frac{1}{3}} \left[1 - \frac{\sigma(3\chi_m - 1)}{9a_{1}}\ln\left(\frac{\tau_{0}}{\tau}\right)\right]^{\frac{1}{4}}.
\label{T_mhd_3}
\end{equation}
For $\tau > \tau_0$, where $\ln(\tau_0/\tau) < 0$, this solution exhibits two distinct thermodynamic behaviors:
When $3\chi_m < 1$ (characteristic of QGP with $\chi_m \sim 0.01$--$0.05$~\cite{Bali:2014kia}), the logarithmic term reduces the temperature, accelerating the cooling rate due to enhanced energy transfer from the fluid to the magnetic field.
Conversely, when $3\chi_m > 1$ (corresponding to hypothetical strong magnetization), the logarithmic term becomes positive, thereby increasing the temperature and slowing the cooling process as the fluid absorbs energy from the magnetic field.

\begin{figure}[tbp!]
\centering
\includegraphics[width=0.9\linewidth]{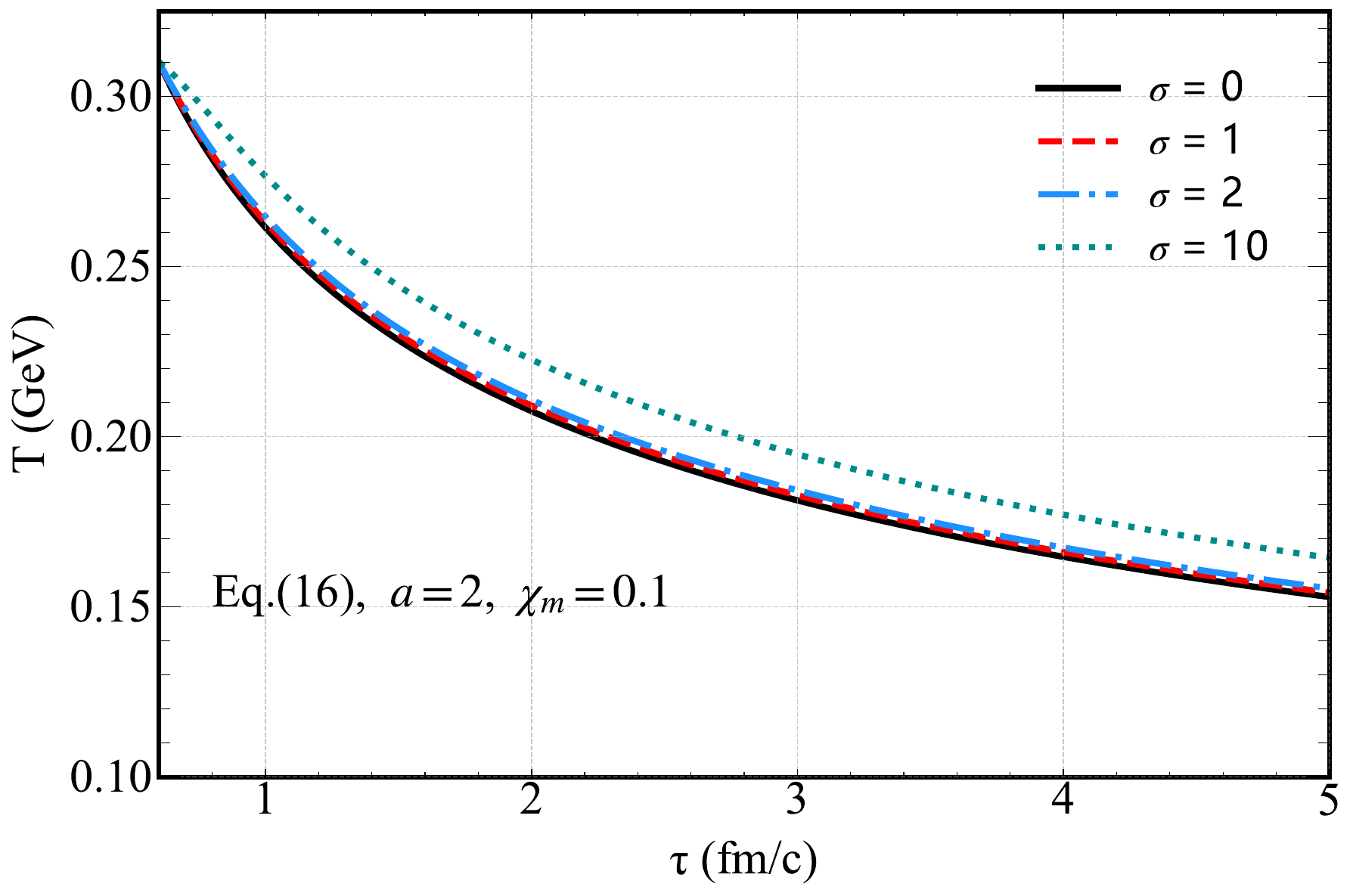} \\
\vspace{0.2cm}
\includegraphics[width=0.9\linewidth]{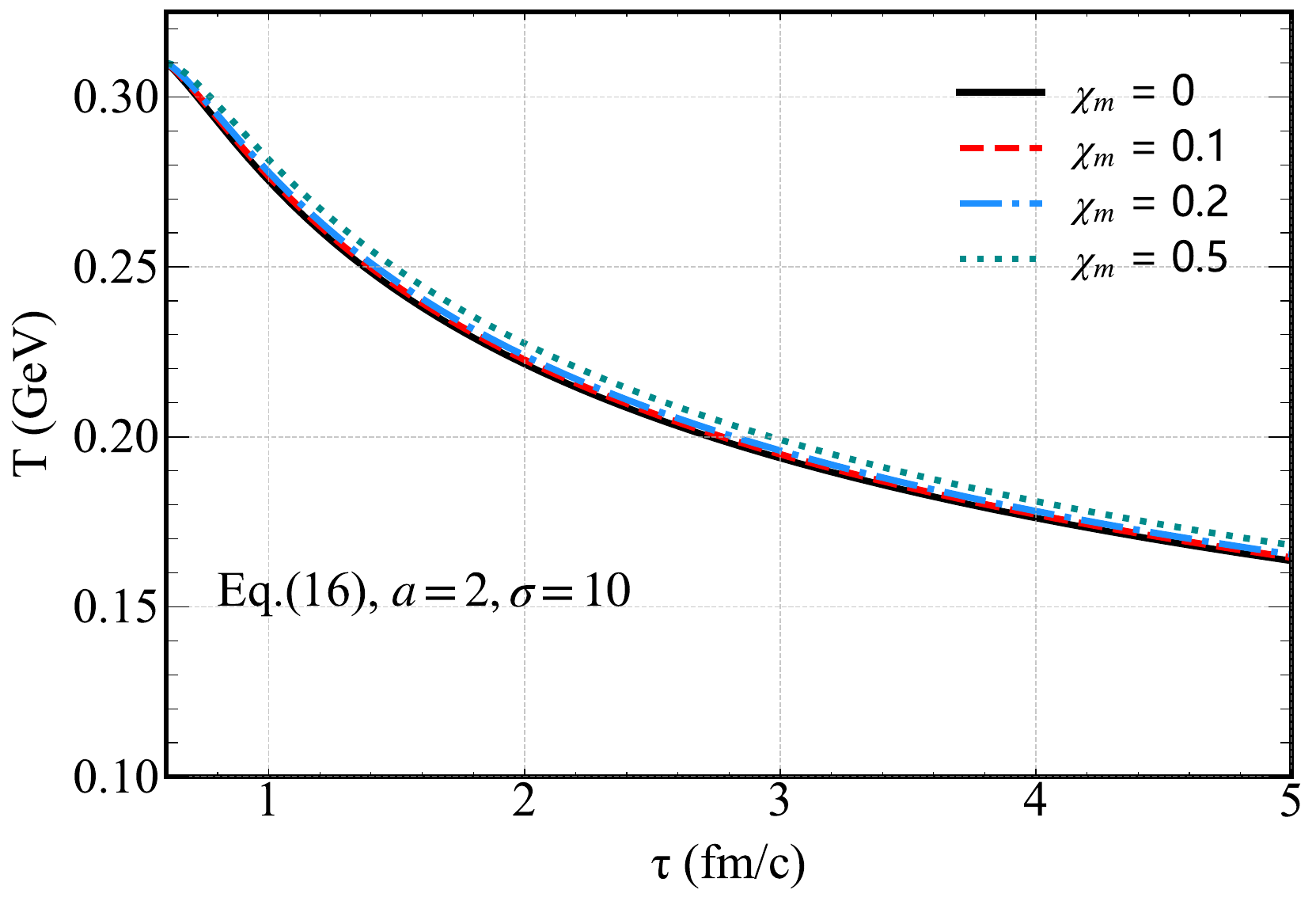}
\caption{(Color online) Evolution of temperature $T$ (Eq.~(\ref{T_mhd_1})) as a function of proper time $\tau$ for different initial magnetic field strengths (upper panel) and constant magnetic susceptibility $\chi_{m}$ (lower panel).}
\label{fig1:a-sigma}
\end{figure}

\begin{figure}[tbp!]
\includegraphics[width=0.9\linewidth]{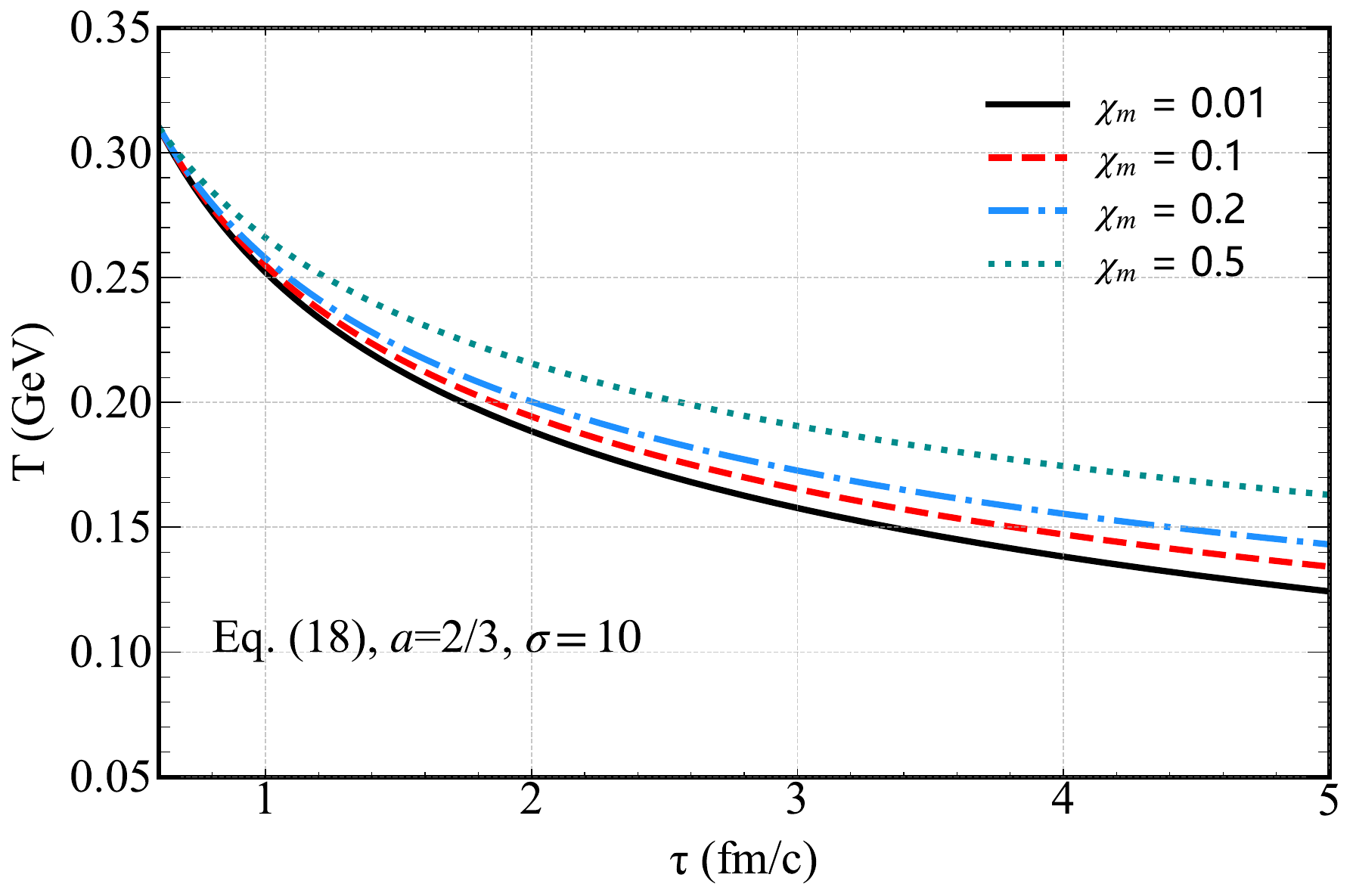} \\
\includegraphics[width=0.9\linewidth]{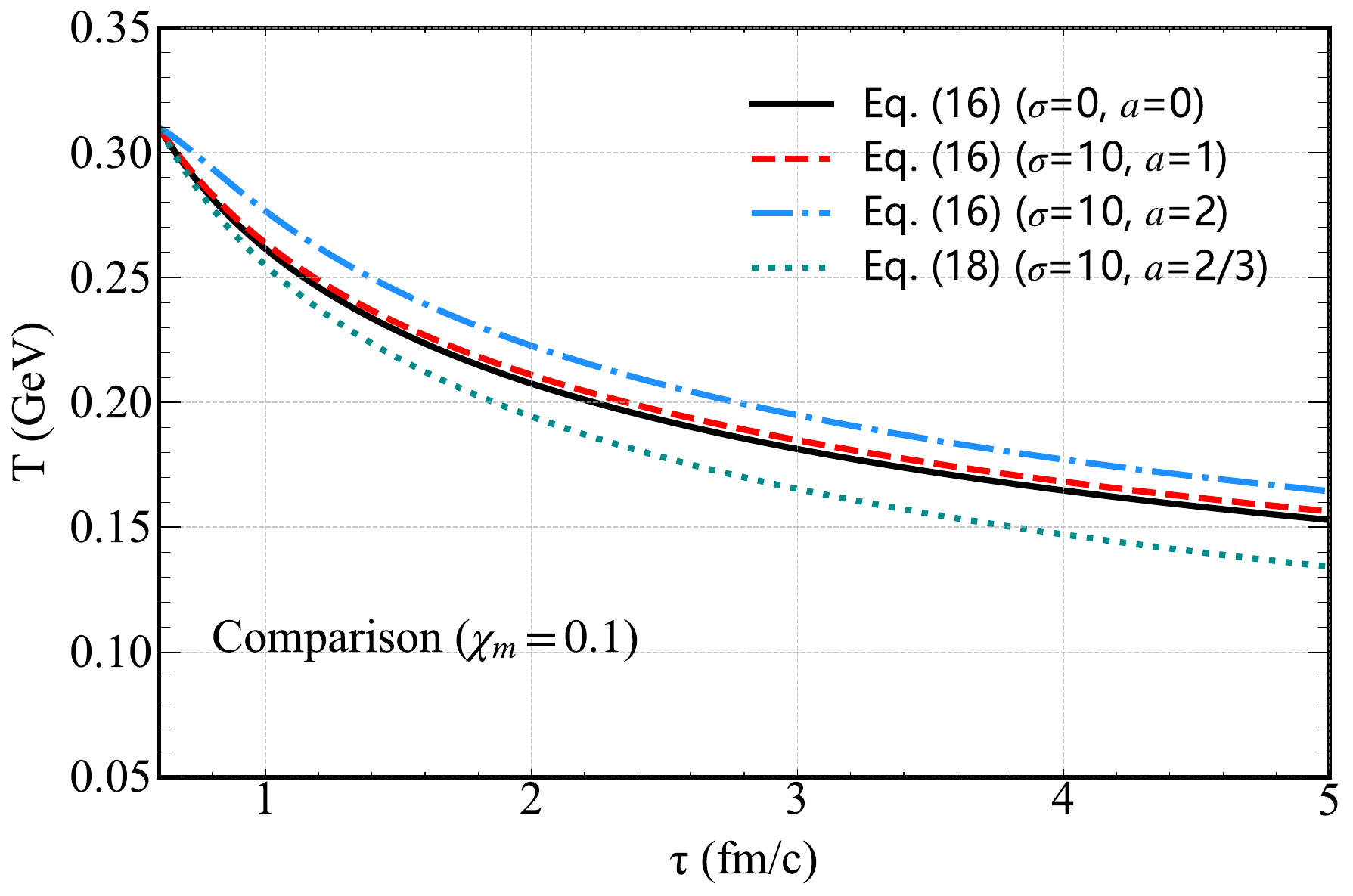}
\caption{(Color online) Evolution of temperature $T$ (Eq.~(\ref{T_mhd_1})) for a magnetic field with as functions of proper time $\tau$ for different initial magnetic field (upper panel) and different magnetic field decay parameter $a$ (lower panel).}
\label{fig2:a-sigma}
\end{figure}

In the upper panel of Fig.~\ref{fig1:a-sigma}, we present the temperature evolution $T$ from Eq.~(\ref{T_mhd_1}) as a function of proper time $\tau$ for $a=2$, fixed $\chi_m=0.1$, and varying $\sigma=0,1,2,10$. We observe that $T$ decays progressively slower with increasing $\sigma$. Since $a=2 > 1-\chi_m$, the magnetic field decays more rapidly than in the ideal MHD case, thereby transferring energy to the QGP. The case $\sigma=0$ (solid line) exhibits the fastest decay, while $\sigma=10$ shows the slowest, reflecting the reheating effect induced by the strong magnetic field~\cite{Pu:2016ayh}.

In the lower panel of Fig.~\ref{fig1:a-sigma}, we display the temperature evolution for $a=2$, fixed $\sigma=10$, and $\chi_m=0,0.1,0.2,0.5$. For $\chi_m>0$ (paramagnetic QGP), the fluid absorbs energy from the magnetic field, thereby retarding the decay compared to the $\chi_m=0$ case (solid line). Larger values of $\chi_m$ strengthen the fluid-field coupling, elevating $T$ at each $\tau$ and further suppressing cooling. This trend confirms that stronger paramagnetism amplifies energy extraction from the magnetic field, thereby mitigating the cooling effect of longitudinal expansion~\cite{Pu:2016ayh}.

In the upper panel of Fig.~\ref{fig2:a-sigma}, we present the temperature evolution $T$ from Eq.~(\ref{T_mhd_3}) as a function of proper time $\tau$ for the critical decay parameter $a=2/3$ and $\chi_m=0.01,0.1,0.2,0.5$. We find that paramagnetic QGP ($\chi_m>0$) enables the fluid to absorb energy from the magnetic field. Larger $\chi_m$ strengthens the fluid-field coupling, thereby suppressing cooling: $T$ decays more slowly with increasing $\chi_m$, with $\chi_m=0.5$ exhibiting the slowest reduction and $\chi_m=0.01$ the fastest.

In the lower panel of Fig.~\ref{fig2:a-sigma}, we compare $T$ as a function of $\tau$ for fixed $\chi_m=0.1$ across four distinct parameter sets from Eqs.~(\ref{T_mhd_1})--(\ref{T_mhd_3}): $\sigma=0$ (vanishing magnetic field), $a=1$ (ideal MHD frozen-flux), $a=2$ (rapid magnetic field decay), and $a=2/3$ (critical decay). The $a=2$ case exhibits the slowest temperature decay and maintains the highest temperature at all $\tau$, while $a=2/3$ corresponds to the fastest decay regime. This trend originates from magneto-fluid energy exchange: larger $a$ accelerates magnetic field decay, transferring more energy to the paramagnetic QGP ($\chi_{m}=0.1$) and thereby suppressing cooling. The $a=2/3$ critical decay case features moderate energy transfer efficiency, resulting in more rapid temperature evolution.

\subsubsection{Temperature-Dependent Magnetic Susceptibility ($\chi_m = \chi_m(T)$)}
For realistic descriptions of the QGP, we adopt temperature-dependent magnetic susceptibilities derived from lattice QCD calculations~\cite{Bali:2014kia,Bali:2020bcn}. The energy conservation equation generalizes to:
\begin{equation}
\frac{\partial T}{\partial \tau} + \frac{(\kappa+1)T}{4\kappa \tau} + \left[1 - a - \chi_{m}(T)\right]\frac{\sigma T_{0}^{4}\tau_{0}^{2a}}{4\kappa a_{1} T^{3} \tau^{2a+1}} = 0.
\label{eq:T-1}
\end{equation}

\begin{figure}[tbp!]
\includegraphics[width=0.9\linewidth]{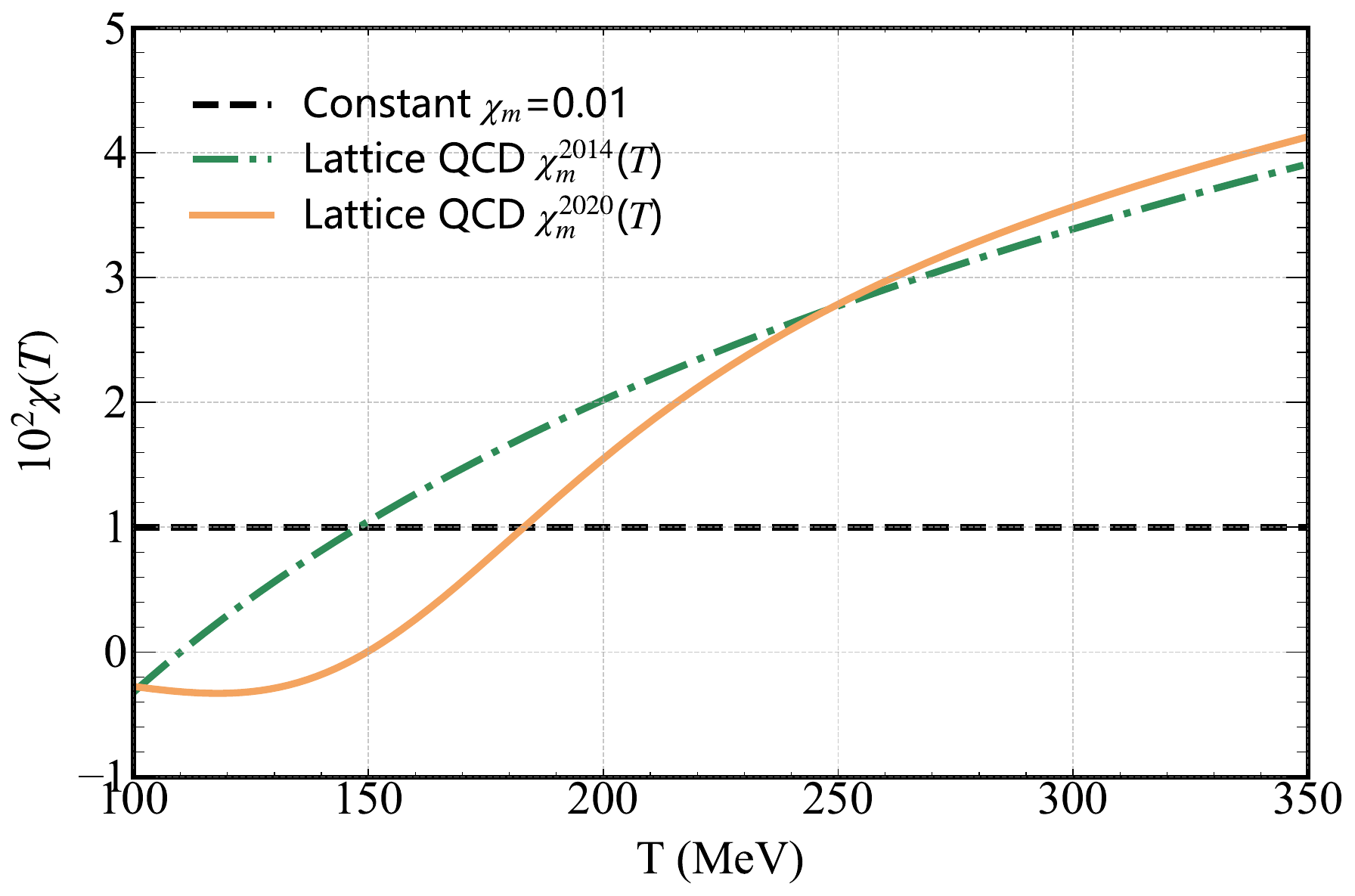} 
\caption{(Color online) Temperature dependence of the magnetic susceptibility $\chi_{m}(T)$ for three prescriptions: constant $\chi_{m}=0.01$, the 2014 lattice QCD result (Eq.~(\ref{eq:chi_2014})), and the 2020 lattice QCD parametrization (Eq.~(\ref{eq:chi_2020})), over 100 $\leq~T~\leq$ 350 MeV.}
\label{f:chim_c}
\end{figure}

We employ two lattice QCD parametrizations for $\chi_m(T)$:

(1) The 2014 lattice result~\cite{Bali:2014kia} for the paramagnetic QGP phase:
\begin{equation}
\chi^{2014}_{m}(T) = \frac{e^2}{3\pi^2} \log\left(\frac{T}{0.11\,\text{GeV}}\right),
\label{eq:chi_2014}
\end{equation}

(2) The 2020 lattice result~\cite{Bali:2020bcn}, describing the temperature-dependent response across the deconfinement transition:
\begin{equation}
\begin{aligned}
\chi_{m}^{2020}(T) &= 2e^{2}\beta_1\log\left(\frac{t}{q_0}\right) \\ 
&\times \frac{1 + g_0/t + g_1/t^2 + g_2/t^3}{1 + g_3/t + g_4/t^2 + g_5/t^3} \exp\left(\frac{-h_3}{t}\right),    
\end{aligned}
\label{eq:chi_2020}
\end{equation}
where $t=T/1$GeV, and the parameters are listed in Table~\ref{tab:parameter}. 

\begin{table}[!h]
\setlength{\tabcolsep}{1.8pt}
\begin{center}
\caption{Parameters for $\chi^{2020}_m(T)$ in Eq.~(\ref{eq:chi_2020}) \cite{Bali:2020bcn}.}
\label{tab:parameter}
\begin{tabular}{|c|c|c|c|c|c|c|c|c|}
\hline
$\beta_1$       & $q_0$  & $g_0$  & $g_1$  & $g_2$ & $g_3$ & $g_4$ & $g_5$ & $h_3$  \\ \hline
$1/(6\pi^2)$    & 0.1497      & 23.99       & -2.085           & 0.1290          & 21.35       & -6.201          & 0.5766          & 0.1544      \\ 
\hline
\end{tabular}
\end{center}
\end{table}

Additionally, for the lattice-QCD-derived $\chi_{m}^{2014}(T)$, an analytical solution is obtained via perturbative methods when $\sigma/a_1 \ll 1$; the detailed derivation and comparison are provided in Appendix~\ref{app:perturbative_mhd}.

In subsequent calculations, we solve Eq.~(\ref{eq:T-1}) numerically by substituting either $\chi^{2014}_m(T)$ or $\chi^{2020}_m(T)$, thereby incorporating the complete temperature-dependent magnetic response of the QGP across both the deconfined and hadronic phases.

In Fig.~\ref{f:chim_c}, we compare the three magnetic susceptibility prescriptions across the temperature range $100$--$350$ MeV, encompassing the hadronic phase, deconfinement region, and hot QGP: constant $\chi_{m}=0.01$ (flat profile), the 2014 lattice QCD result (monotonic logarithmic growth with $T$), and the 2020 lattice QCD parametrization (realistic temperature dependence deviating from the 2014 form at low $T$ and smoothing around $T_c$). These differences in $\chi_{m}(T)$ modify the QGP-magnetic field energy exchange, leading to distinct temperature evolution histories and thermal photon yields.

\begin{figure}[tbp!]
\includegraphics[width=0.9\linewidth]{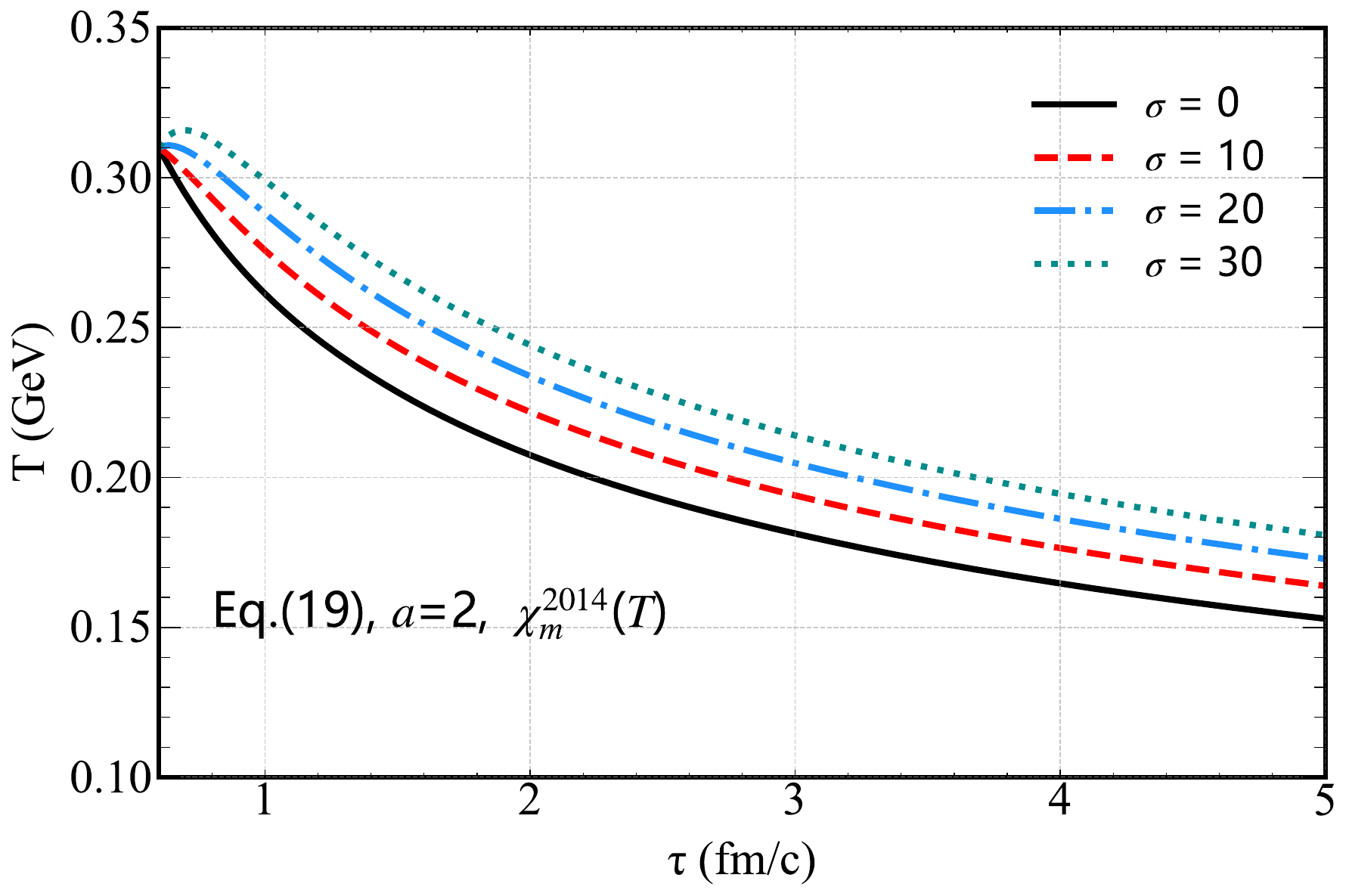} \\
\includegraphics[width=0.9\linewidth]{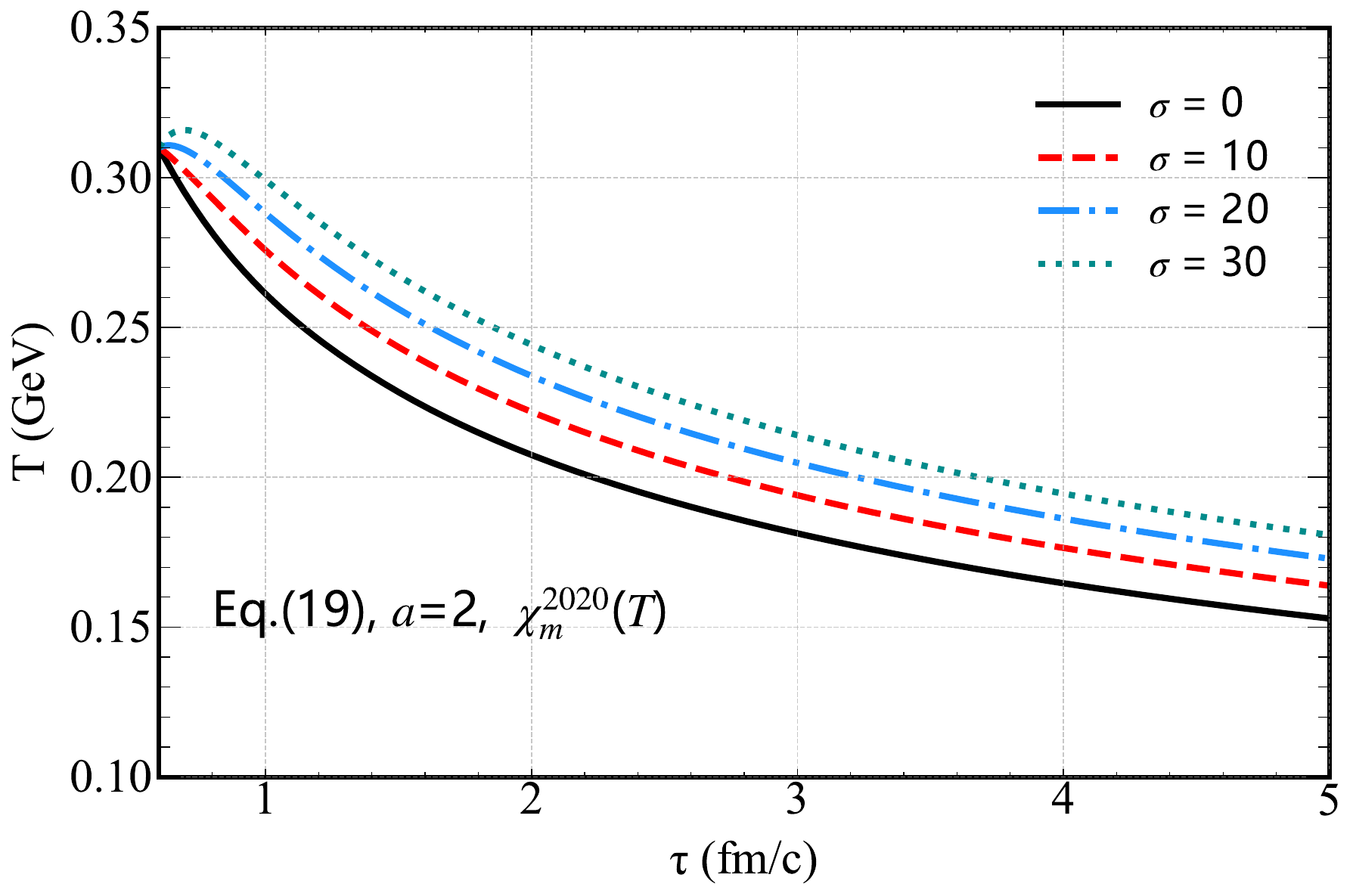}
\caption{(Color online) Temperature $T$ as a function of proper time $\tau$ for $a=2$, temperature-dependent magnetic susceptibility, and varying initial magnetic field strength $\sigma=0,10,20,30$ (Eq.~(\ref{eq:mhd-0})). Upper panel: $\chi_m(T)_{2014}$ (Eq.~(\ref{eq:chi_2014})). Lower panel: $\chi_m(T)_{2020}$ (Eq.~(\ref{eq:chi_2020})).}
\label{fig:chi2014-2020}
\end{figure}

In the upper panel of Fig.~\ref{fig:chi2014-2020}, we present the temperature $T$ as a function of proper time $\tau$ for $a=2$, the temperature-dependent $\chi^{2014}_m(T)$ from Eq.~(\ref{eq:chi_2014}), and initial magnetic field strengths $\sigma=0,10,20,30$. Consistent with the energy conservation equation, larger $\sigma$ (corresponding to stronger initial magnetic fields) intensifies energy transfer from the decaying magnetic field to the paramagnetic QGP, thereby suppressing the decay of $T$. The $\sigma=0$ case (absence of magnetic field effects) exhibits the most rapid cooling, while $\sigma=30$ sustains the highest temperature at all $\tau$---consistent with the logarithmic growth of $\chi^{2014}_m(T)$, which enhances fluid-field coupling.

In the lower panel of Fig.~\ref{fig:chi2014-2020}, we display the temperature evolution for $a=2$, $\chi_m^{2020}(T)$ from Eq.~(\ref{eq:chi_2020}), and the same set of $\sigma$ values. The evolutionary trend is analogous to that observed with $\chi^{2014}_m(T)$: the decay of $T$ slows progressively with increasing $\sigma$, driven by energy transfer from the decaying magnetic field to the QGP. Although $\chi^{2020}_m(T)$ exhibits deviations from $\chi^{2014}_m(T)$ in the low-temperature regime (Fig.~\ref{f:chim_c}), the overall dependence of the temperature evolution on $\sigma$ persists. This reflects the dominant role of the initial magnetic field strength in regulating QGP cooling during rapid magnetic field decay ($a=2$).

\begin{figure}[tbp!]
\includegraphics[width=0.9\linewidth]{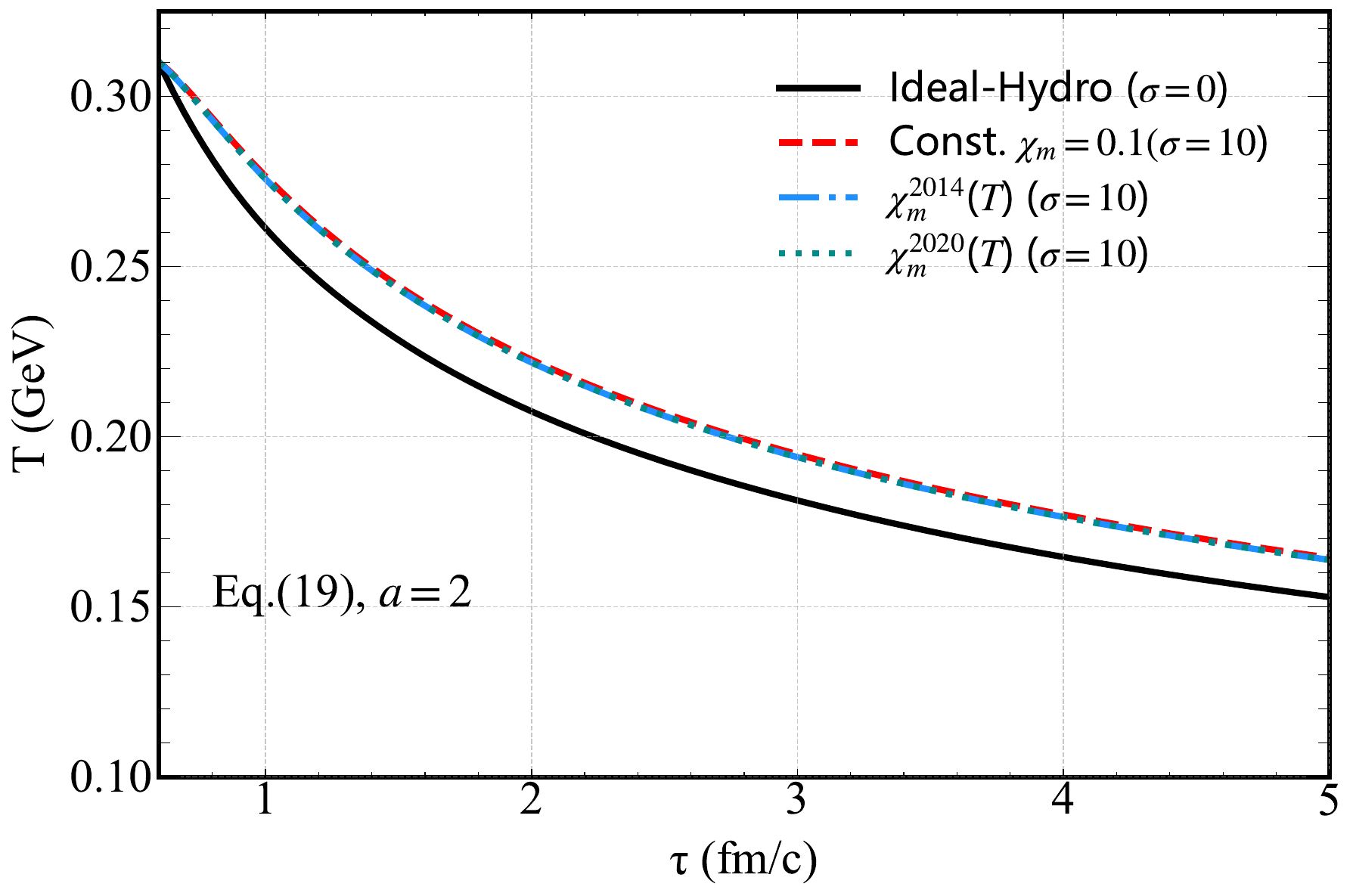} 
\caption{(Color online) Evolution of temperature $T$ (Eq.~(\ref{eq:T-1})) as functions of proper time $\tau$ for different $\chi_{m}$.}
\label{fig:chiall}
\end{figure}

Finally, in Fig.~\ref{fig:chiall}, we compare the temperature evolution for the three $\chi_m$ prescriptions (constant $\chi_m=0.1$, $\chi_m^{2014}(T)$, and $\chi_m^{2020}(T)$). We find that all cases with $\sigma=10$ exhibit slower temperature decay and higher temperatures than ideal hydrodynamics ($\sigma=0$), as the paramagnetic QGP absorbs energy from the decaying magnetic field. Notably, the three $\chi_m$ prescriptions exhibit consistent decay trends---this originates from the key factor $(1-a-\chi_m) = -1-\chi_m$ for $a=2$. In the QGP temperature regime ($150$--$350$ MeV), $\chi_m$ for all prescriptions lies within a narrow range ($0.01$--$0.2$), leading to negligible variation in $(1-a-\chi_m)$ and comparable magneto-fluid energy exchange efficiency. Subtle differences in the absolute temperature arise from their distinct temperature-dependent coupling strengths, resulting in similar effects of the three magnetic susceptibility prescriptions on photon emission yields.

\subsection{Thermal Photons Rate}
\label{sec:2-C}
Thermal photons emitted during the quark-gluon plasma (QGP) phase stand as pristine probes of the medium's thermodynamic state and dynamical evolution. Unlike hadrons, these photons escape the QGP without undergoing strong interactions, preserving unaltered information about their production conditions-from temperature and density to magnetic-field effects.

Photon production in the QGP arises from key quantum chromodynamic (QCD) processes, with both leading-order and higher-order contributions playing distinct roles. The core channels include quark/gluon Compton scattering ($q(\bar{q})g \to q(\bar{q})\gamma$) and quark-antiquark annihilation ($q\bar{q} \to g\gamma$), which dominate the baseline photon yield \cite{Traxler:1995kx,Steffen:2001pv}. Notably, higher-order processes are non-negligible: two-loop bremsstrahlung (photon emission from charged partons accelerated by medium interactions) contributes comparably to leading-order channels \cite{Steffen:2001pv,Bhatt:2010cy}, while off-shell $q\bar{q}$ annihilation with additional parton scatterings becomes significant in the dense, high-temperature QGP environment. Modern descriptions of photon production thus require incorporating these higher-order effects to align with experimental data \cite{Traxler:1995kx,Bhatt:2010cy}.

To quantify thermal photon yields, we employ one-loop perturbative QCD (pQCD) calculations combined with hard thermal loop (HTL) resummation-an essential technique for accounting for medium effects (e.g., color charge screening) in the QGP \cite{Traxler:1995kx}. Below, we present the production rates for the dominant processes:

\subsubsection{Compton Scattering + $q\bar{q}$ Annihilation (C+A)}
This composite channel unifies contributions from quark-gluon Compton scattering and quark-antiquark annihilation, described by the HTL-resummed rate:
\begin{equation}
E\frac{dN_{\textrm{C+A}}}{d^{3}pd^{4}x} = \frac{1}{2\pi^{2}}\alpha\alpha_{s}\left(\sum_{f}e_{f}^{2}\right)T^{2}e^{-E/T}\ln\left(\frac{cE}{\alpha_{s}T}\right),
\label{eq:ca}
\end{equation}
where $\alpha = 1/137$ (electromagnetic fine-structure constant), $\alpha_s$ (strong coupling constant), and $e_f$ (electric charge of quark flavor $f$, in units of $e$) are standard pQCD parameters. We consider light quarks ($u, d$) with $e_u = 2/3$ and $e_d = -1/3$, while $c \approx 0.23$ is an HTL-derived constant. The exponential $e^{-E/T}$ encodes the Boltzmann distribution of thermal partons, and the $\alpha_s(T)$ follows the temperature-dependent parametrization \cite{Karsch:1987kz}:
\begin{equation}
\alpha_{s}(T) = \frac{6\pi}{(33-2N_f)\ln(8T/T_c)},
\end{equation}
with $N_f = 2$ (active light quark flavors) and $T_c = 0.14$ GeV (QGP-hadron phase transition temperature).

\subsubsection{Bremsstrahlung (Bre)}
Bremsstrahlung describes photon emission from quarks scattered by other partons (quarks/gluons) in the QGP, with a rate dominated by soft photon production:
\begin{equation}
E\frac{dN_{\textrm{Bre}}}{d^{3}pd^{4}x} = \frac{1}{8\pi^{5}}\alpha\alpha_{s}\left(\sum_{f}e_{f}^{2}\right)\frac{T^{4}}{E^{2}}e^{-E/T}(J_T - J_L)I(E,T),
\label{eq:bre}
\end{equation}
Here, $J_T \approx 1.11$ and $J_L \approx -1.06$ are transverse/longitudinal flow parameters from two-loop calculations \cite{Srivastava:1999ekv,Steffen:2001pv}, and $I(E,T)$ encapsulates kinematic effects via polylogarithmic terms:
\begin{equation}
\begin{aligned}
I(E,T) &= 3\zeta(3) + \frac{\pi^2}{6}\frac{E}{T} + \frac{E^2}{T^2}\ln 2 + 4\textrm{Li}_3(-e^{-|E|/T}) \\
&+ \frac{2E}{T}\textrm{Li}_2(-e^{-|E|/T}) - \left(\frac{E}{T}\right)^2\ln\left(1 + e^{-|E|/T}\right),
\end{aligned}
\end{equation}
where $\zeta(3) \approx 1.202$ (Riemann zeta function) and $\textrm{Li}_a(z)$ (polylogarithm) account for thermal parton phase space integrals. The $\propto 1/E^2$ dependence highlights the soft nature of this channel, making it dominant at low photon energies.

\subsubsection{$q\bar{q}$-Annihilation with Additional Scattering (A+S)}
This higher-order process involves $q\bar{q}$ annihilation coupled to a secondary medium scattering event, enhancing photon production at intermediate-to-high energies. Its rate is:
\begin{equation}
E\frac{dN_{\textrm{A+S}}}{d^{3}pd^{4}x} = \frac{8}{3\pi^{5}}\alpha\alpha_{s}\left(\sum_{f}e_{f}^{2}\right)ETe^{-E/T}(J_T - J_L),
\label{eq:as}
\end{equation}
In contrast to bremsstrahlung, this channel scales linearly with photon energy $E$, rendering it more prominent than the latter at higher $E$ while retaining the thermal Boltzmann factor $e^{-E/T}$.


\begin{figure}[tbp!]
\includegraphics[width=0.9\linewidth]{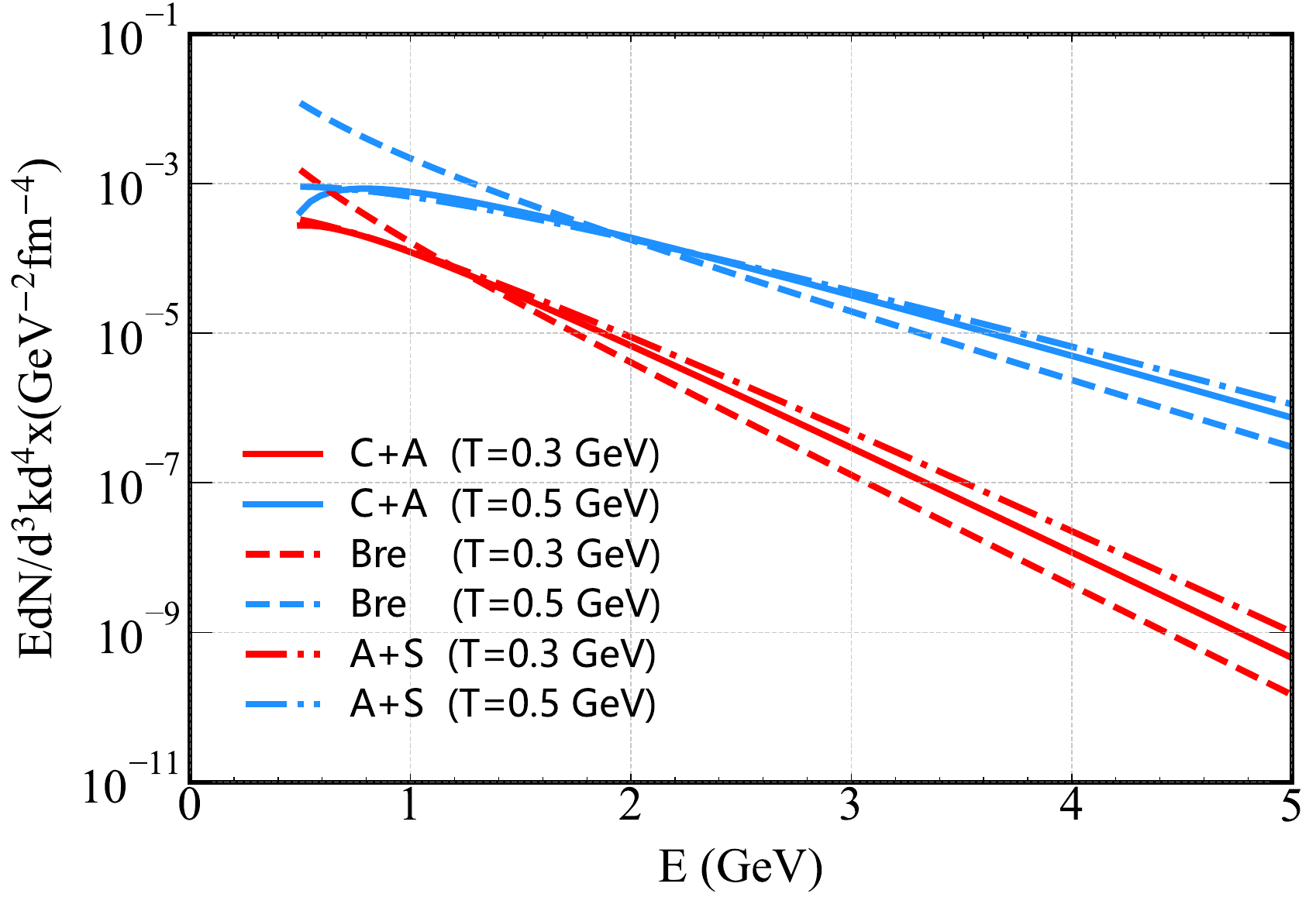}
\caption{(Color online) Thermal photon production rates $E\frac{dN}{d^3kd^4x}$ as a function of photon energy $E$ for C+A, Bremsstrahlung (Bre), and A+S channels at $T=0.3$ GeV and $T=0.5$ GeV.}
\label{f:etchi}
\end{figure}


In Fig.~\ref{f:etchi}, we present the thermal photon production rates $E\frac{dN}{d^3kd^4x}$ for the three key channels (C+A, Bremsstrahlung (Bre), A+S) as a function of photon energy $E$, calculated at two fixed QGP temperatures: $T=0.3$ GeV and $T=0.5$ GeV. Bremsstrahlung channel dominates the low-energy regime: it is the primary contributor up to $E \sim 1.2$ GeV (for $T=0.3$ GeV) and $E \sim 1.8$ GeV (for $T=0.5$ GeV), while C+A and A+S processes take over at higher energies. This energy-dependent hierarchy arises from their distinct kinematic dependences: Bremsstrahlung is suppressed at high $E$ by the $1/E^2$ factor, whereas C+A (logarithmic $E$ dependence) and A+S (linear $E$ dependence) gain strength with increasing $E$. Additionally, all three production rates are notably enhanced at the hotter $T=0.5$ GeV relative to $T=0.3$ GeV-an effect driven by elevated thermal parton occupation numbers and more frequent parton collisions in the hotter QGP, which is consistent with prior theoretical investigations \cite{Steffen:2001pv,Bhatt:2010cy}.

The total thermal photon rate is the sum of these individual contributions:
\begin{equation}
E\frac{dN_{\textrm{total}}}{d^{3}pd^{4}x} = E\frac{dN_{\textrm{C+A}}}{d^{3}pd^{4}x} + E\frac{dN_{\textrm{Bre}}}{d^{3}pd^{4}x} + E\frac{dN_{\textrm{A+S}}}{d^{3}pd^{4}x}.
\label{f:cabsa-total}
\end{equation}

\begin{figure}[tbp!]
\includegraphics[width=0.9\linewidth]{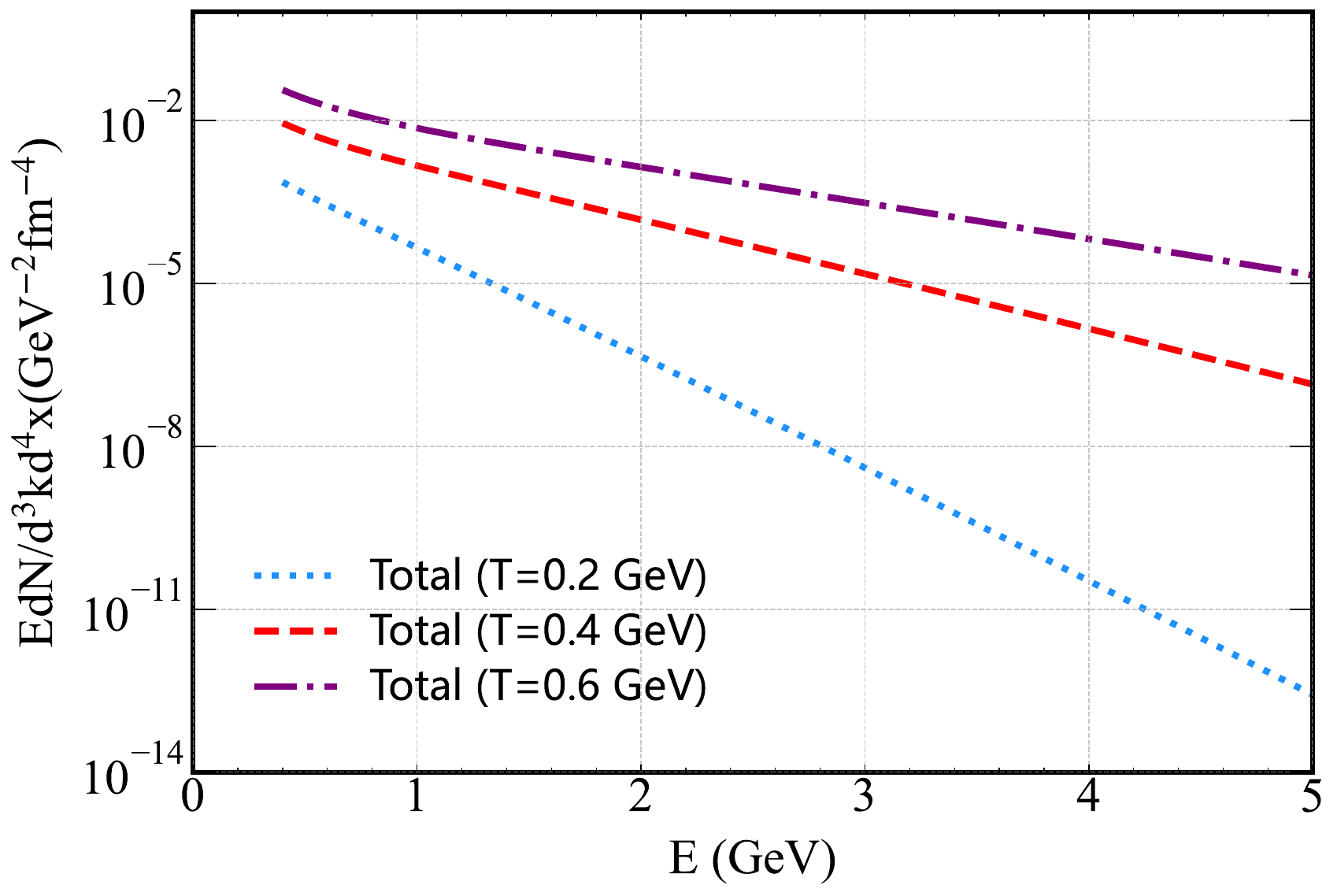}
\caption{(Color online)  Total thermal photon production rates as a function of photon energy $E$ for $T=0.2$ GeV, $T=0.4$ GeV, and $T=0.6$ GeV.}
\label{f:etchi-2}
\end{figure}

In Fig.~\ref{f:etchi-2}, we plot this total rate for $T=0.2$ GeV, $T=0.4$ GeV, and $T=0.6$ GeV, emphasizing the strong temperature dependence of photon production: at $E \sim 3$ GeV, the rate for $T=0.6$ GeV is approximately $10^5$ times larger than that for $T=0.2$ GeV. This enhancement underscores the sensitivity of thermal photon spectra to the QGP's temperature evolution a key motivation for using photons to probe the medium's thermal history.

To connect these differential rates to observable spectra, we integrate the total rate over the QGP's spacetime evolution, following the approach in Refs.~\cite{Traxler:1995kx,Steffen:2001pv,Bhatt:2010cy}:
\begin{equation}
\left(\frac{dN}{d^{2}\pt dy}\right)_{y,~\pt} = Q_c \int_{\tau_0}^{\tau_f} d\tau\,\tau \int_{-y_{\textrm{nuc}}}^{y_{\textrm{nuc}}} d\eta_s \left(E\frac{dN_{\textrm{total}}}{d^{3}pd^{4}x}\right),
\label{eq:photon_int}
\end{equation}
where $\tau_0 = 0.5$ fm/c (initial proper time), when the QGP cools to $T_c=140$ MeV). The integral over $\tau$ accounts for photon production throughout the QGP's lifetime, weighted by the expansion factor $\tau$ (a consequence of longitudinal boost invariance), while the integral over $\eta_s$ (space-time rapidity) spans the nuclear rapidity range $y_{\textrm{nuc}}$. The factor $Q_c \sim 180$ fm$^2$ is the nuclear cross section for Au+Au collisions~\cite{Srivastava:1999ekv,Traxler:1995kx,Steffen:2001pv,Bhatt:2010cy}.

Within the Pu-Bjorken flow framework (neglecting transverse expansion), the fluid four-velocity is $u^\mu = (\cosh\eta_s, 0, 0, \sinh\eta_s)$, and the photon energy in the QGP rest frame is $E = \pt \cosh(y - \eta_s)$, where $\pt$ is the transverse momentum and $y$ is the photon rapidity. This relation links the photon's observed momentum to its energy in the medium, enabling the transformation from theoretical rates to realistic observables.

\subsection{Weak magnetic field corrections to the distribution functions}
\label{sec:2-C}

We follow Refs.~\cite{Sun:2023pil,Sun:2023rhh,Sun:2024isb} to calculate the weak magnetic field corrections ($f_{\textrm{EM}}$) to the distribution functions. In prior hydrodynamics-based studies, dissipative effects in the QGP medium have been incorporated into thermal photon production calculations; magnetic-field-induced modifications to photon emission rates have also been investigated~\cite{Hattori:2016cnt}. These effects are introduced via viscous corrections to the quark and gluon distribution functions, $f = f_0 + \delta f$, where $f_0$ denotes the equilibrium Boltzmann distribution and $\delta f$ represents the out-of-equilibrium correction, linear in either shear or bulk viscosity. Analogously, a weak external electromagnetic field induces an additional correction to the quark distribution function, which can be written as $f_q = \bar{f} + f_{\textrm{EM}}$~\cite{Sun:2023pil}, with $\bar{f}$ being the distribution function in the absence of electromagnetic fields and $f_{\textrm{EM}}$ the electromagnetic-field-induced correction.

At leading order in $|eB|/T^2$, and for a locally charge-neutral system (e.g., the QGP created in high-energy heavy-ion collisions), the out-of-equilibrium effects driven by a weak external electromagnetic field can be solved via the Boltzmann-Vlasov equation. Adopting the relaxation time approximation, the equation of motion for the quark distribution function reads:
\begin{equation}
p^\mu \partial_\mu f + e_f F^{\mu\nu} p_\mu \frac{\partial f}{\partial p^\nu} = -\frac{p \cdot u}{\tau_R} \delta f,
\end{equation}
where $e_f$ is the electric charge of the quark flavor $f$ (with $e_u = 2e/3$ and $e_d = -e/3$), $\tau_R$ is the parton relaxation time (characterizing medium interactions and related to transport coefficients), and $\delta f = f - f_0$ is the total out-of-equilibrium correction.

This equation can be solved analytically using the Chapman-Enskog method. Expanding $\delta f$ in powers of $eF^{\mu\nu}/T^2$, the leading-order solution for the electromagnetic correction $f_{\textrm{EM}}$ (isolating the field-induced contribution) is~\cite{Sun:2023pil}:
\begin{equation}
f_{\textrm{EM}} = e_f F^{\mu\nu} p_\mu u_\nu \frac{\tau_R}{p \cdot u} \frac{n_{\textrm{eq}}(1 - n_{\textrm{eq}})}{T},
\end{equation}
where $n_{\textrm{eq}} = f_0$ (the equilibrium number density) and the relation $-\partial n_{\textrm{eq}}/\partial p^\nu = u_\nu n_{\textrm{eq}}(1 - n_{\textrm{eq}})/T$ has been used. To relate $\tau_R$ to a physically measurable quantity, one can utilize the definition of the electrical current $j^\mu = \sigma_{\textrm{el}} E^\mu$. 
Substituting $f_{\textrm{EM}}$ into current definition yields $\tau_R \propto \sigma_{\textrm{el}} / T^2$~\cite{Sun:2023pil}. Replacing $\tau_R$ with $\sigma_{\textrm{el}}$ in the expression for $f_{\textrm{EM}}$, one obtains~\cite{Sun:2023pil}:
\begin{equation}
f_{\textrm{EM}} = \frac{c}{8\alpha_{\textrm{EM}}} \frac{\sigma_{\textrm{el}} n_{\textrm{eq}}(1 - n_{\textrm{eq}})}{T^3 p \cdot u} e_f F^{\mu\nu} p_\mu p_\nu,
\label{eq:em_c}
\end{equation}
where $\alpha_{\textrm{EM}} = 1/137$ is the electromagnetic fine-structure constant, and $c$ is a constant dependent on the number of active quark flavors: $c = 9\pi/10$ for a 2-flavor (quark-antiquark) massless system with classical statistics, and $c = 3\pi/4$ for 3-flavor quarks.
While Eq.~(\ref{eq:em_c}) applies generally to weak electromagnetic fields, we focus on the transverse magnetic field component $B_y$ (relevant to heavy-ion collision geometries) following Refs.~\cite{Deng:2012pc,Roy:2015kma}. For QGP temperatures above the crossover temperature $T_c$, the weak field condition $|eB|/m_\pi^2 \ll 1$ ($\sigma$ is not very large) ensures $f_{\textrm{EM}}/n_{\textrm{eq}} \ll 1$, justifying the perturbative expansion.

Under Bjorken flow conditions (longitudinal boost invariance, neglecting transverse expansion), we simplify $f_{\textrm{EM}}$ for a weak magnetic field along the $y$-direction. Using the fluid four-velocity $u^\mu = (\cosh\eta_s, 0, 0, \sinh\eta_s)$ (with $\eta_s$ the spacetime rapidity) and the relation between photon energy in the QGP rest frame and lab frame ($E = p_T \cosh(y - \eta_s)$, where $p_T$ is the transverse momentum and $y$ is photon rapidity), the electromagnetic correction reduces to:
\begin{equation}
f_{\textrm{EM}} = e_f B \frac{c}{8\alpha_{\textrm{EM}}} \frac{\sigma_{\textrm{el}} n_{\textrm{eq}}}{T^3} \frac{\sinh\eta_s}{\cosh(y - \eta_s)}.
\end{equation}
Incorporating the time evolution of the magnetic field, $B(\tau) = \sigma T_0^2 (\tau_0/\tau)^{2a}$ (with $\sigma$ a constant, $T_0$ the initial QGP temperature, $\tau_0 = 0.5$ fm/c the initial proper time, and $a$ the magnetic field decay exponent), the full quark distribution function including the weak magnetic field correction is:
\begin{equation}
\begin{aligned}
f &= f_0 \left(1 + e_f \sigma T_0^2 \left(\frac{\tau_0}{\tau}\right)^{2a} \frac{c}{8\alpha_{\textrm{EM}}} \frac{\sigma_{\textrm{el}}}{T^3} \frac{\sinh\eta_s}{\cosh(y - \eta_s)}\right) \\
&= f_0 \left(1 + \delta f_{\textrm{EM}}\right),
\end{aligned}
\end{equation}
where $\delta f_{\textrm{EM}}$ denotes the dimensionless electromagnetic correction factor.

We now describe precisely how $f_{\rm EM}$ enters the photon production calculation. In the small-angle (soft) approximation, the photon emission rate from $2\to2$ scattering processes is proportional to the phase-space distribution of the incoming quarks and antiquarks~\cite{Sun:2023pil,Sun:2023rhh}, namely $E\, d\mathcal{R}/d^3p \propto \mathcal{I}\,(f_q + f_{\bar q})$, where the medium-dependent conversion factor $\mathcal{I} = \int d^3p/(2\pi)^3 E_p\,(f_g+f_q+f_{\bar q})$ characterizes the quark-to-photon transition rate inside the QGP~\cite{Blaizot:2014jna}. Substituting the full quark distribution $f_q = f_0 + f_{\rm EM}$ (and analogously $f_{\bar q} = f_0 + f_{\bar{\rm EM}}$, with $f_{\bar{\rm EM}} = -f_{\rm EM}$ since the antiquark carries opposite electric charge $e_{\bar q} = -e_q$), the photon emission rate separates additively:
\begin{equation}
E\frac{dN_{\rm total}}{d^3p\,d^4x} = E\frac{d\bar{N}}{d^3p\,d^4x} \;+\; E\frac{dN_{\rm EM}}{d^3p\,d^4x},
\label{eq:rate_additive}
\end{equation}
where $E d\bar{N}/d^3p d^4x$ is the background rate given by the sum of the C+A, Bremsstrahlung, and A+S channels (Eqs.~\ref{eq:ca}, \ref{eq:bre}, \ref{eq:as}), each evaluated with the equilibrium distribution $f_0$. The magnetic-field-induced correction is
\begin{equation}
E\frac{dN_{\rm EM}}{d^3p\,d^4x} \propto \mathcal{I}\,\sum_f \bigl(f_{\rm EM}^{(f)} + f_{\rm EM}^{(\bar f)}\bigr),
\label{eq:rate_em}
\end{equation}
where the sum runs over the active quark flavors ($u$ and $d$). Although $f_{\rm EM}^{\bar q} = -f_{\rm EM}^{q}$ at the level of the distribution function, the net EM contribution to the photon rate does not vanish after phase-space integration, for the following reasons. First, the kinematic factor $F^{\mu\nu}p_\mu u_\nu$ in $f_{\rm EM}$ weights different momentum regions of quarks and antiquarks differently. Second, and more importantly, the rapidity-odd structure $f_{\rm EM} \propto \sinh\eta_s$, when convoluted with the rapidity-odd dipole moment inherent in the tilted fireball geometry~\cite{Chatterjee:2017ahy}, yields a non-zero contribution after the spacetime integration over a symmetric rapidity window $y\in[-y_{\rm nuc}, y_{\rm nuc}]$ (cf.\ Eq.~\ref{eq:photon_int}). At the practical level, $f_{\rm EM}$ is evaluated point-by-point on the spacetime grid and integrated together with the conversion factor $\mathcal{I}$ following the identical numerical quadrature as for the equilibrium contribution. The additive decomposition in Eq.~\ref{eq:rate_additive} is justified by the weak-field condition $|eB|/T^2 \ll 1$, which guarantees $|f_{\rm EM}| \ll f_0$ throughout the QGP evolution and validates the linear perturbation expansion.

It is worth emphasizing that our treatment of the magnetic-field correction as an additive term in the photon rate (Eq.~\ref{eq:rate_additive}) is the natural extension of the well-established procedure for incorporating viscous corrections in photon production~\cite{Paquet:2015lta,Gale:2021emg}. There, the shear and bulk viscous corrections $\delta f_\pi$ and $\delta f_\Pi$ enter the rate analogously through $f = f_0 + \delta f_\pi + \delta f_\Pi$, and the corresponding corrections to the photon spectrum are obtained by substituting the full distribution function into the phase-space integral. Our implementation of $f_{\rm EM}$ follows precisely this logic.

\vspace{4pt}
We note the limitations of the present framework: (1) we adopt ideal MHD evolution and neglect dissipative corrections ($\delta f_{\rm vis}$) to the distribution function from shear/bulk viscosity \cite{Bhatt:2010cy,Sun:2023pil}, which may modify photon production rates at late stages of the QGP evolution; (2) thermal photon production in the hot hadron gas (HHG) phase is not included, despite its contribution to the low-$p_T$ spectra after QGP freeze-out \cite{Steffen:2001pv,Shen:2013cca,Chatterjee:2017akg}; (3) the present treatment of the $f_{\rm EM}$ correction in the photon rate adopts the small-angle approximation, and a full phase-space integration of the exact $2\to2$ matrix elements with the modified quark distribution $f_0+f_{\rm EM}$ would be desirable for a more quantitative analysis. These effects, alongside strong magnetic-field-induced modifications to parton interaction vertices, will be explored in future work to develop a more comprehensive description of thermal photon production in heavy-ion collisions.

\begin{figure}[tbp!]
\includegraphics[width=0.9\linewidth]{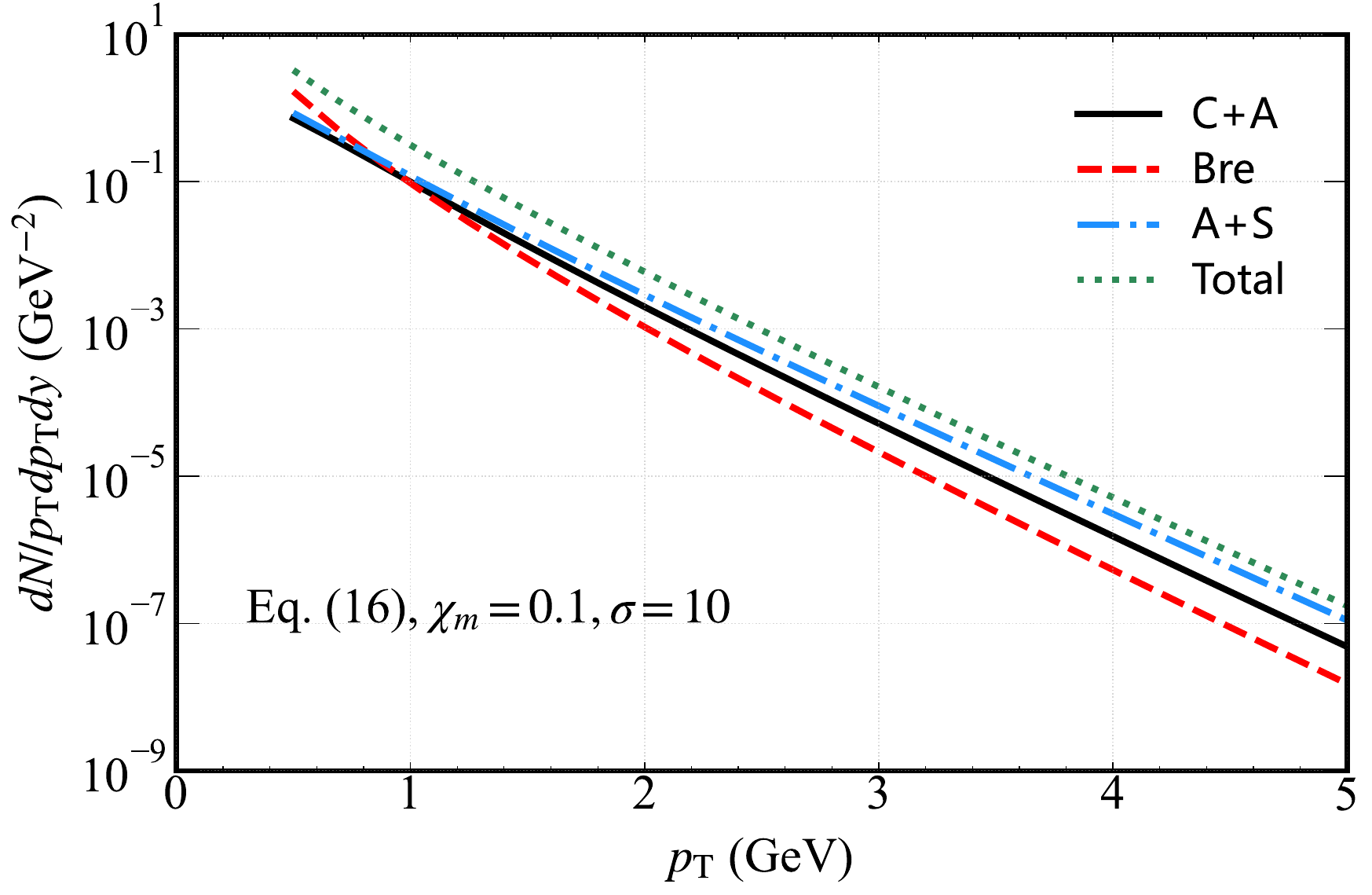}
\caption{(Color online) Hard thermal photon spectrum as a function of transverse momentum $p_T$ for the C+A, Bremsstrahlung (Bre), A+S channels, and total yield, calculated with $\chi_m=0.1$, $\sigma=10$, and MHD solution.}
\label{fig8:diff}
\end{figure}

\section{Results and discussion}  
\label{section-4}

This section presents our central results regarding the sensitivity of thermal photon spectra to the external magnetic field $B$, the QGP magnetic susceptibility $\chi_m$, and the magnetic field-induced correction $f_{\rm EM}$.

We compute the temperature profiles using the MHD solution (Eq.~(\ref{T_mhd_1})) with benchmark parameters: magnetic field decay exponent $a=2$~\cite{Deng:2012pc,Pang:2016igs}, initial field strength $\sigma=10$~\cite{Roy:2015kma}, and susceptibility $\chi_m=0.1$. The photon production rates for the dominant channels---Compton scattering with $q\bar{q}$ annihilation (C+A), bremsstrahlung (Bre), and $q\bar{q}$ annihilation with additional scattering (A+S)---are evaluated via Eqs.~(\ref{eq:ca}), (\ref{eq:bre}), and (\ref{eq:as}). The final spectra are obtained through spacetime integration of these rates using Eq.~(\ref{eq:photon_int}), accounting for emission throughout the QGP lifetime. Following previous studies~\cite{Traxler:1995kx,Steffen:2001pv,Bhatt:2010cy}, we focus on the midrapidity region ($y=0$) with RHIC-relevant parameters: initial temperature $T_0 = 0.31~\mathrm{GeV}$, freeze-out temperature $T_f = 0.14~\mathrm{GeV}$, initial proper time $\tau_0 = 0.5~\mathrm{fm}/c$, freeze-out proper time $\tau_f \approx 5.5\pm2.0~\mathrm{fm}/c$, and nuclear rapidity $y_{\mathrm{nuc}} = 5.3$.


Fig.~\ref{fig8:diff} displays the $p_T$-differential thermal photon spectrum for $\chi_m=0.1$ with channel decompositions. The spectrum exhibits three distinct regimes: bremsstrahlung dominates at low $p_T$ ($<0.8~\mathrm{GeV}$), consistent with its characteristic $\propto E^{-2}$ soft-photon enhancement; the C+A and A+S channels prevail above $0.8~\mathrm{GeV}$; while above $2~\mathrm{GeV}$, the total spectrum converges to the A+S distribution due to its linear $p_T$ dependence. The $\chi_m$ dependence manifests globally through modified temperature evolution: by regulating magnetic-field-QGP energy exchange, $\chi_m$ reshapes the spacetime temperature profile of the QGP, thereby modulating photon yields across all channels and $p_T$ regimes. The magnetic-field-induced correction $f_{\rm EM}$ is also included.

\begin{figure}[tbp!]
\includegraphics[width=0.85\linewidth]{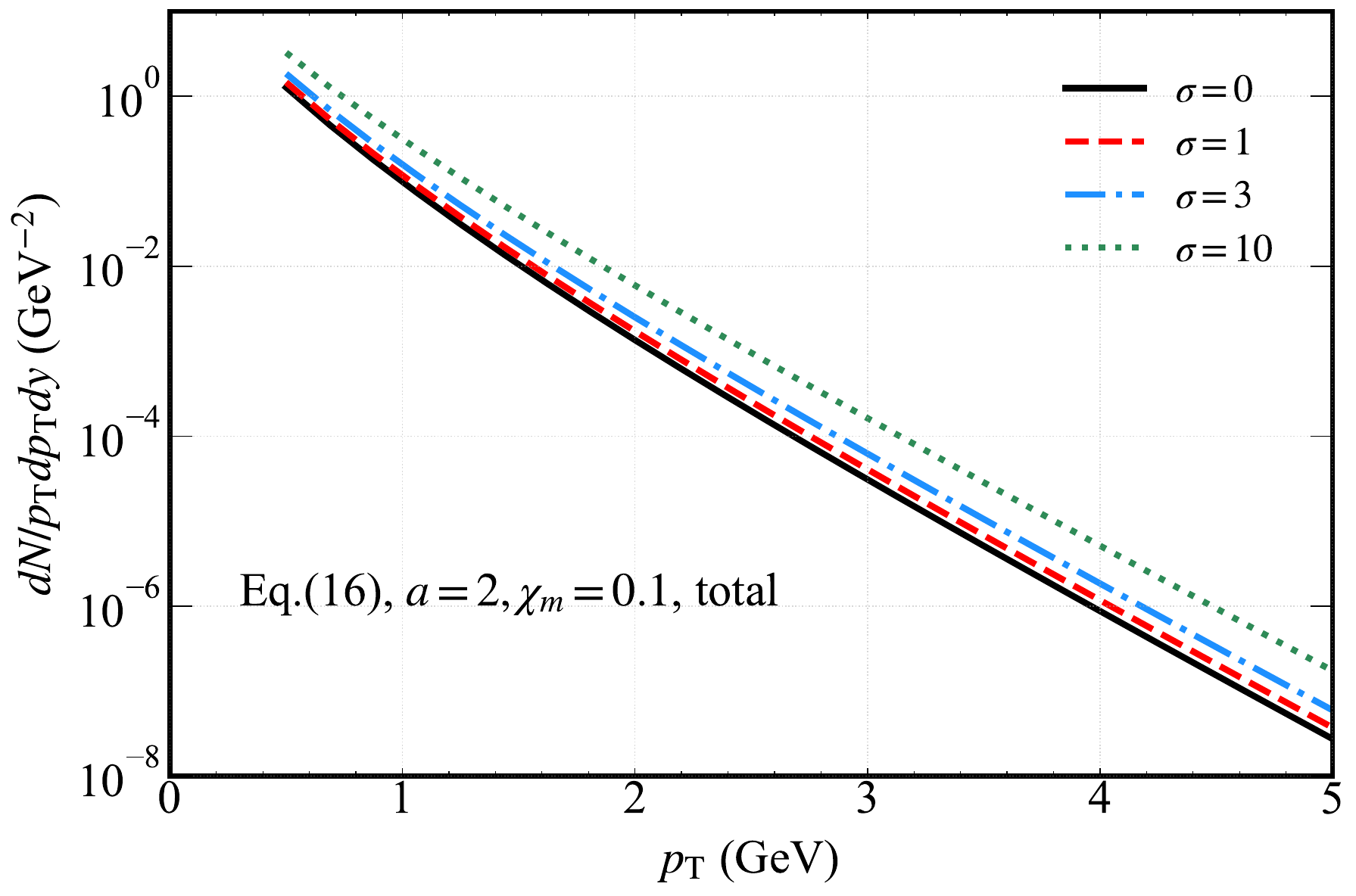}\\
\includegraphics[width=0.85\linewidth]{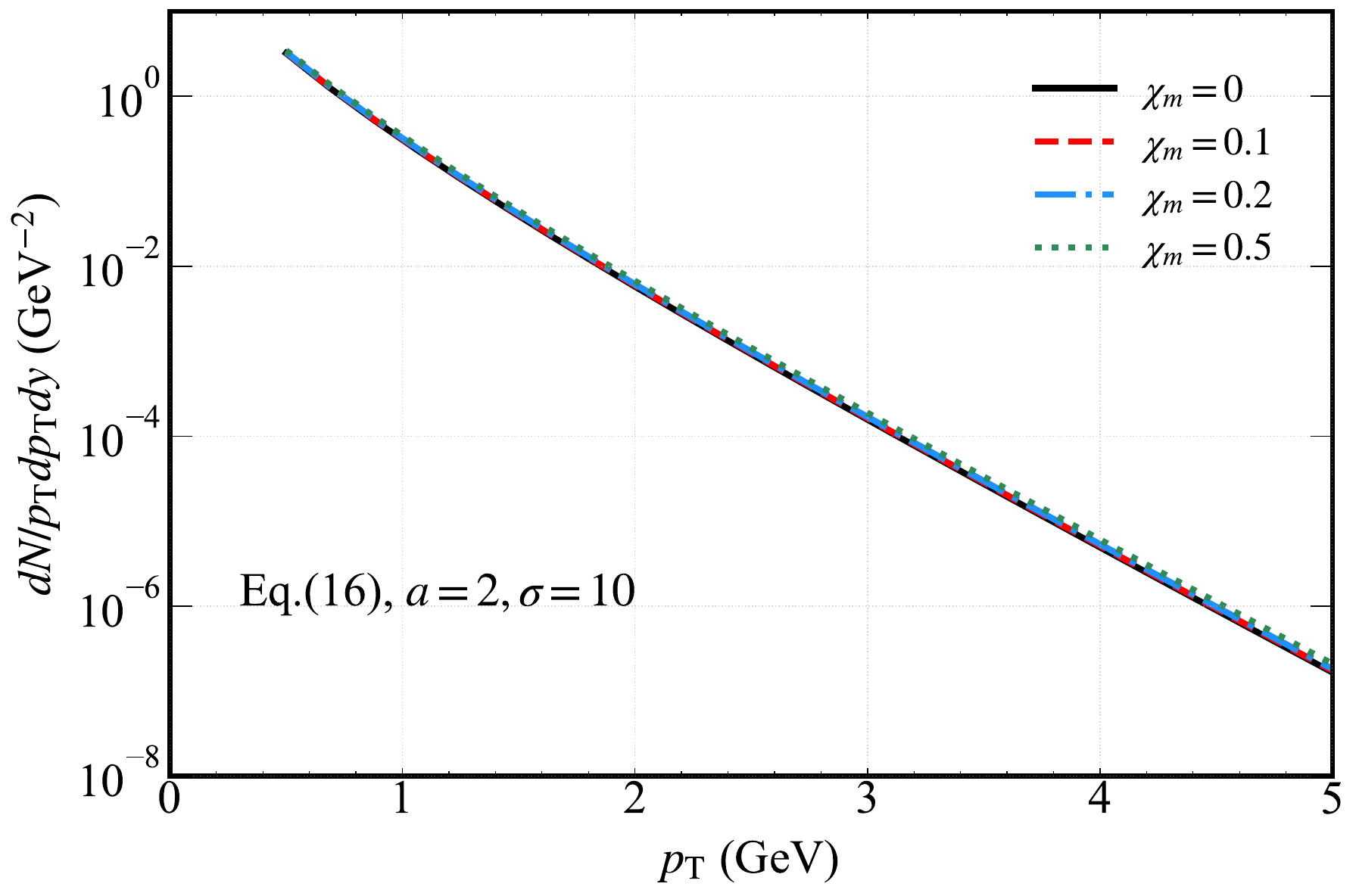}
\caption{(Color online) Total thermal photon production spectra as a function of transverse momentum \(p_T\). Upper panel: Results for different initial magnetic field strengths \(\sigma=0,~1,~3,~10\) with \(a=2\) and \(\chi_m=0.1\). Lower panel: Results for different magnetic susceptibilities \(\chi_m=0,~0.1,~0.2,~0.5\) with \(a=2\) and \(\sigma=10\). All calculations are based on the MHD solution Eq.~(\ref{T_mhd_1}) and spacetime integration Eq.~(\ref{eq:photon_int}).}
\label{f:fig8_chi}
\end{figure}

In the upper panel of Fig.~\ref{f:fig8_chi}, we present the total thermal photon spectra for different initial magnetic field strengths \(\sigma=0,~1,~3,~10\), with temperature profiles derived from Eq.~(\ref{T_mhd_1}). We observe that stronger \(\sigma\) (e.g., \(\sigma=10\)) significantly enhances the photon rate across all \(p_T\), as a more intense initial magnetic field transfers more energy to the QGP via non-zero \(\chi_m\), thereby slowing the temperature decline and extending the emission duration. In contrast, weak \(\sigma\) (e.g., \(\sigma=1\)) has negligible impact: the spectrum is nearly identical to \(\sigma=0\) (field-free case), as the weak field cannot substantially perturb the QGP thermal dynamics. This confirms that the initial magnetic field strength, in combination with \(\chi_m\), plays a crucial role in regulating thermal photon emission.

In the lower panel of Fig.~\ref{f:fig8_chi}, we display the total thermal photon production spectra for different QGP magnetic susceptibilities \(\chi_m=0,~0.1,~0.2,~0.5\), with temperature profiles derived from the MHD solution Eq.~(\ref{T_mhd_1}). Notably, increasing \(\chi_m\) does not lead to significant enhancement of the photon rate across the entire \(p_T\) range; instead, the spectra for all \(\chi_m\) values show negligible differences, with no obvious hierarchy between \(\chi_m=0.5\) and \(\chi_m=0\). This behavior originates from the limited variation of the fluid-field coupling strength: for \(a=2\), the coupled term \((1-a-\chi_m)\) in the energy conservation equation simplifies to \(-1-\chi_m\). Within the explored \(\chi_m\) range (\(0 \leq \chi_m \leq 0.5\)), this term only varies between \(-1\) and \(-1.5\), resulting in minimal differences in the efficiency of magnetic field-QGP energy exchange. Consequently, the QGP temperature decay profiles differ only slightly across different \(\chi_m\), leading to nearly identical effective emission time windows and thermal parton occupation numbers. As a result, the total thermal photon production remains largely unchanged. 

\begin{figure}[tbp!]
\includegraphics[width=0.85\linewidth]{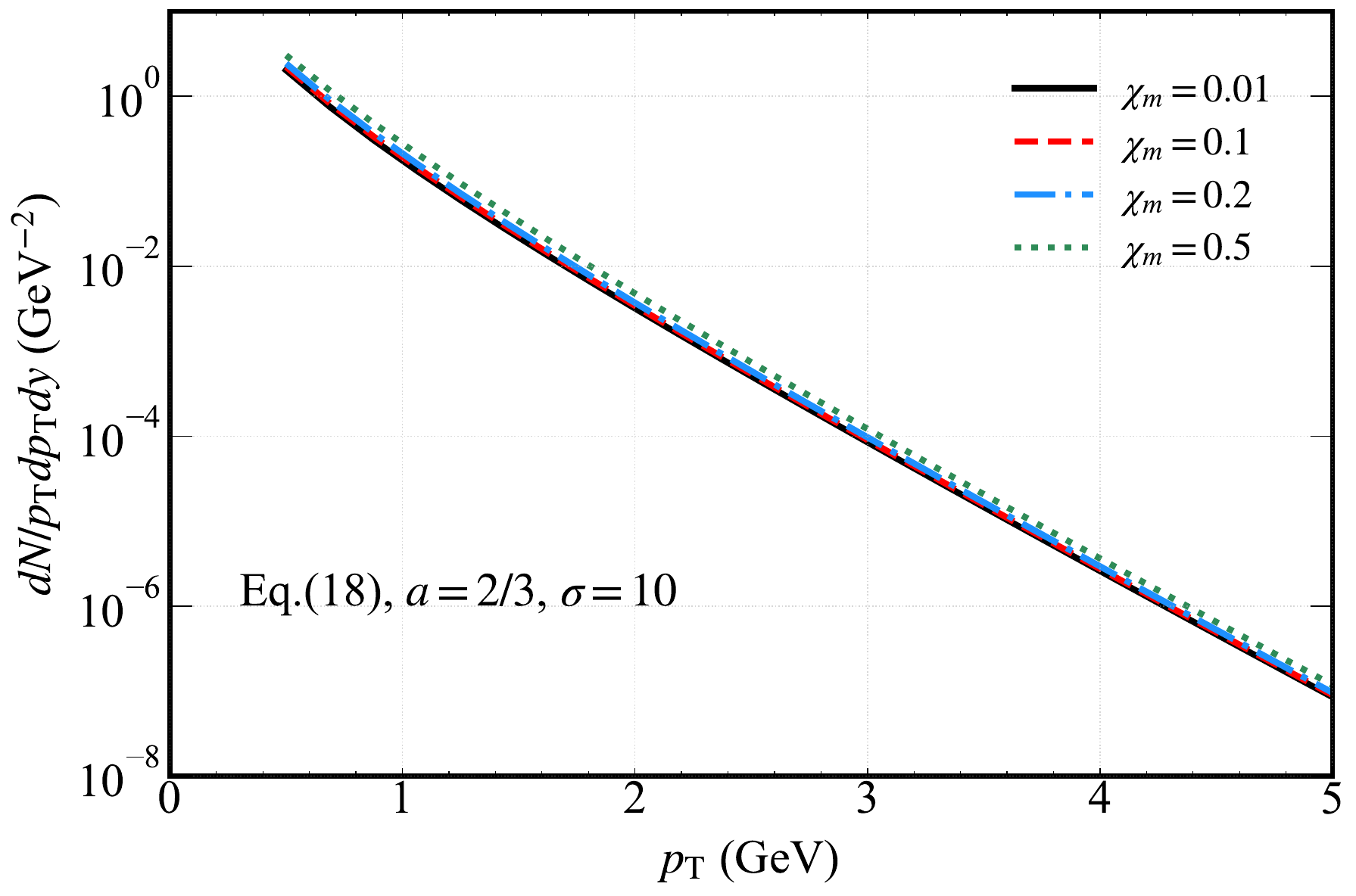}\\
\includegraphics[width=0.85\linewidth]{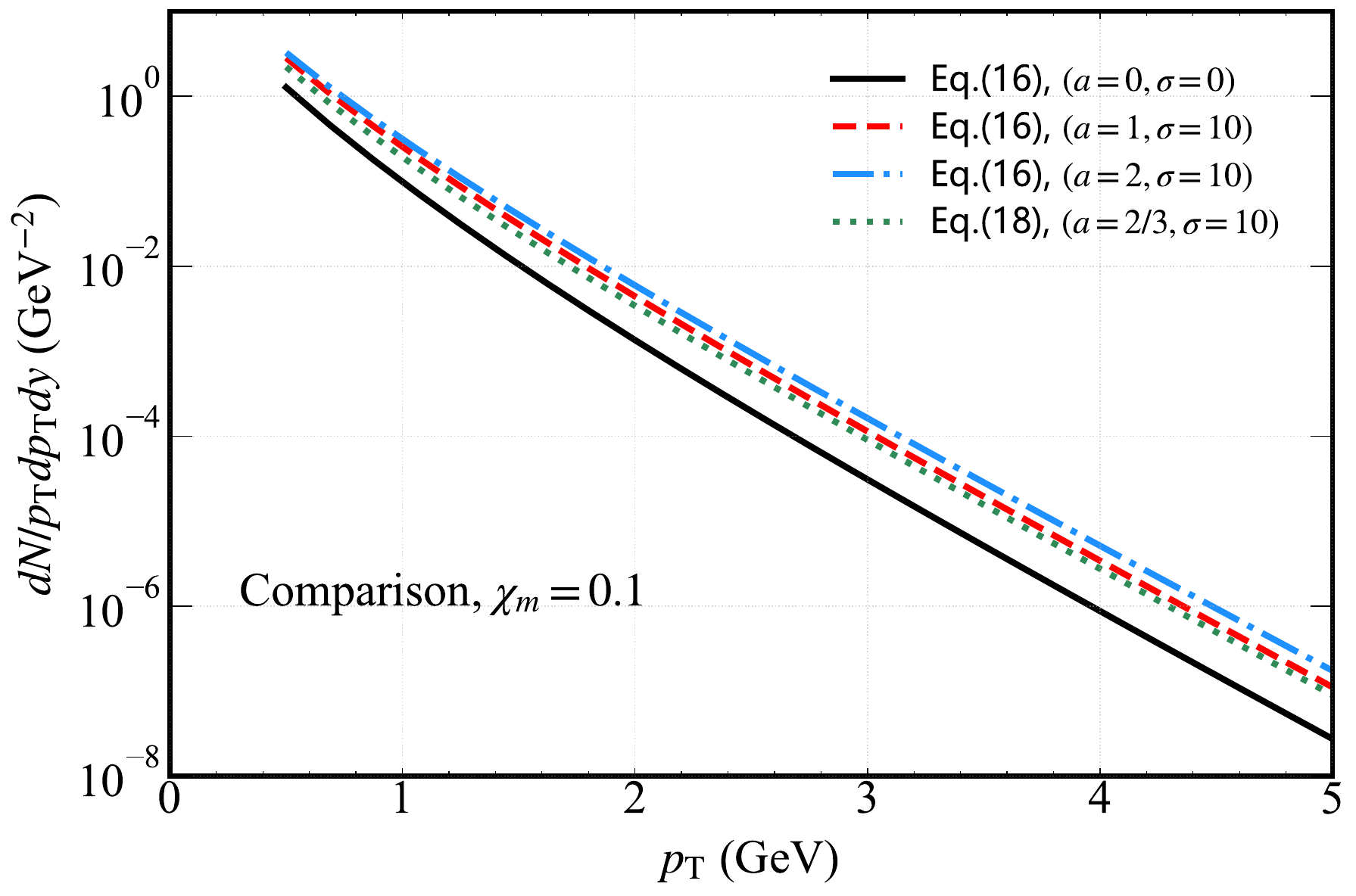}
\caption{(Color online) Total thermal photon production spectra as a function of transverse momentum. Upper panel: Results for \(\chi_m=0.01,~0.1,~0.2,~0.5\) with \(a=2/3\) and \(\sigma=10\). Lower panel: Comparison of cases with fixed \(\chi_m=0.1\).}
\label{f:fig9_chi}
\end{figure}

In the upper panel of Fig.~\ref{f:fig9_chi}, we present the total thermal photon spectra for \(\chi_m=0.01,~0.1,~0.2,~0.5\) (with fixed \(a=2/3\), \(\sigma=10\); MHD solution Eq.~(\ref{T_mhd_3})). Notably, all spectra show negligible differences across the entire \(p_T\) range; increasing \(\chi_m\) does not induce significant photon yield modification. This arises from the coupled term \((1-a-\chi_m)=1/3-\chi_m\), which varies only between \(\sim0.32\) (\(\chi_m=0.01\)) and \(\sim-0.17\) (\(\chi_m=0.5\)), leading to similar magnetic field-QGP energy exchange efficiency and thermal evolution, thus resulting in consistent photon yields.

In the lower panel of Fig.~\ref{f:fig9_chi}, we compare the total thermal photon spectra with fixed \(\chi_m=0.1\), using Eqs.~(\ref{T_mhd_1}) and (\ref{T_mhd_3}). The ideal-hydro case ($\sigma=0$) shows the lowest yield across all \(p_T\) due to the absence of magnetic field energy input to the QGP. All magnetized cases (\(\sigma=10\)) exhibit enhanced yields, with smaller \(a\) leading to higher rates: Eq.~(\ref{T_mhd_3}) (\(a=2/3\)) and Eq.~(\ref{T_mhd_1}) (\(a=1\)) yield comparable spectra, both higher than Eq.~(\ref{T_mhd_1}) (\(a=2\)). This occurs because slower decay prolongs field-QGP energy exchange (mediated by \(\chi_m=0.1\)), slowing QGP cooling and extending the photon emission window. The consistency between MHD prescriptions confirms that magnetic field persistence (governed by \(a\)) is the core driver of enhancement, underscoring \(a\) as the dominant regulator---finite \(\sigma\) provides the energy source, while smaller \(a\) maximizes QGP energy utilization for observable yield boosts.

\begin{figure}[tbp!]
\includegraphics[width=0.85\linewidth]{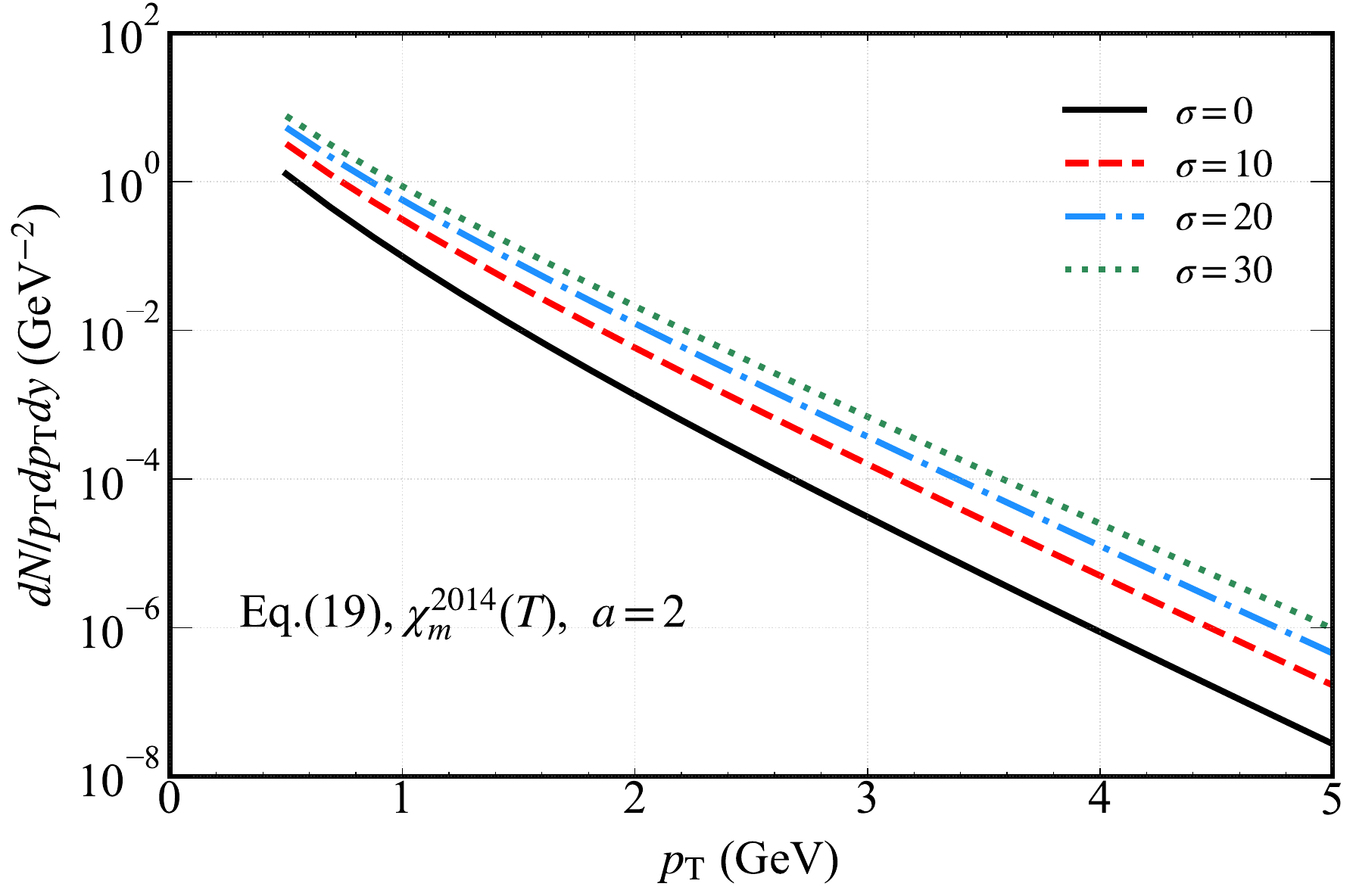}\\
\includegraphics[width=0.85\linewidth]{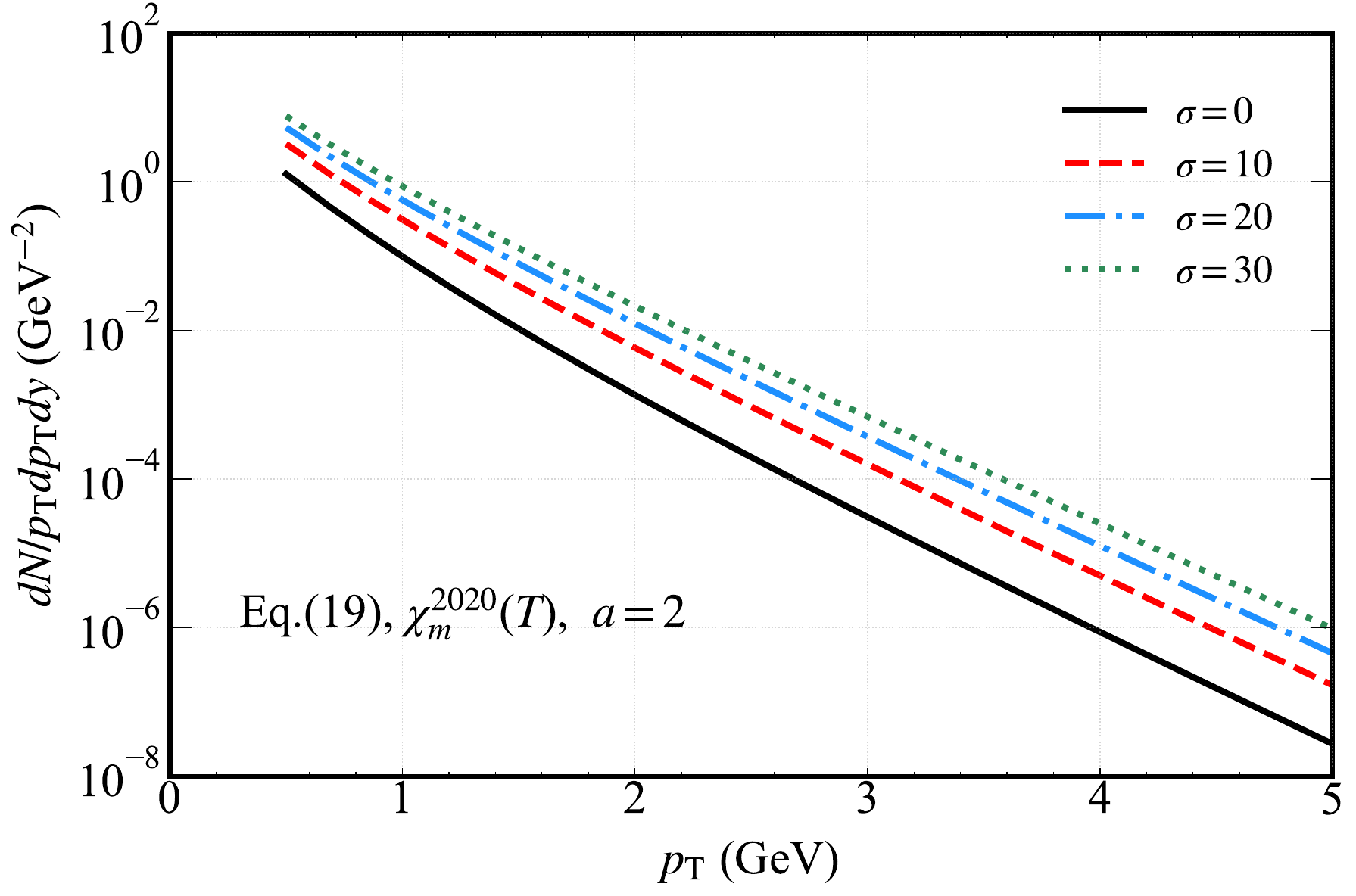}
\caption{(Color online) Total thermal photon production spectra as a function of transverse momentum. Upper panel: Results with $\chi_{m}^{2014}$($T$) for \(\sigma=0,~10,~20,~30\) and \(a=2\). Lower panel: Results with $\chi_{m}^{2020}$($T$) for the same \(\sigma\) and \(a=2\).}
\label{f:fig10_chi}
\end{figure}

In the upper panel of Fig.~\ref{f:fig10_chi}, we examine the role of initial magnetic field strength \(\sigma\) in thermal photon production, adopting the 2014 lattice-QCD-derived temperature-dependent susceptibility $\chi_{m}^{2014}$($T$). For \(\sigma=0,~10,~20,~30\), the spectra exhibit a clear monotonic trend: the field-free case (\(\sigma=0\)) delivers the lowest yield across all \(p_T\), while \(\sigma=30\) (strong initial field) achieves the most significant enhancement, with \(\sigma=10\) and \(20\) following sequentially. This behavior reflects the physics of magnetic field-QGP energy transfer: stronger \(\sigma\) injects more energy into the medium through magnetic susceptibility, slowing thermal cooling and extending the effective window for parton collisions.

In the lower panel of Fig.~\ref{f:fig10_chi}, using $\chi_{m}^{2020}$($T$) with the same \(\sigma\) range, we replicate the core physics of the upper panel: \(\sigma\) remains the key driver of yield enhancement, with \(\sigma=0\), \(10\), \(20\) and \(30\) ordered across all \(p_T\). Minor quantitative offsets emerge at intermediate \(p_T\) (e.g., \(1\sim2\) GeV), attributable to the distinct temperature dependences of $\chi_{m}^{2014}$($T$) and $\chi_{m}^{2020}$($T$) from lattice calculations. Critically, these differences do not alter the fundamental conclusion: the choice of \(\chi_m(T)\) prescription only induces negligible modifications, while initial magnetic field strength \(\sigma\) dominates the thermal photon yield enhancement.

\begin{figure}[tbp!]
\includegraphics[width=0.85\linewidth]{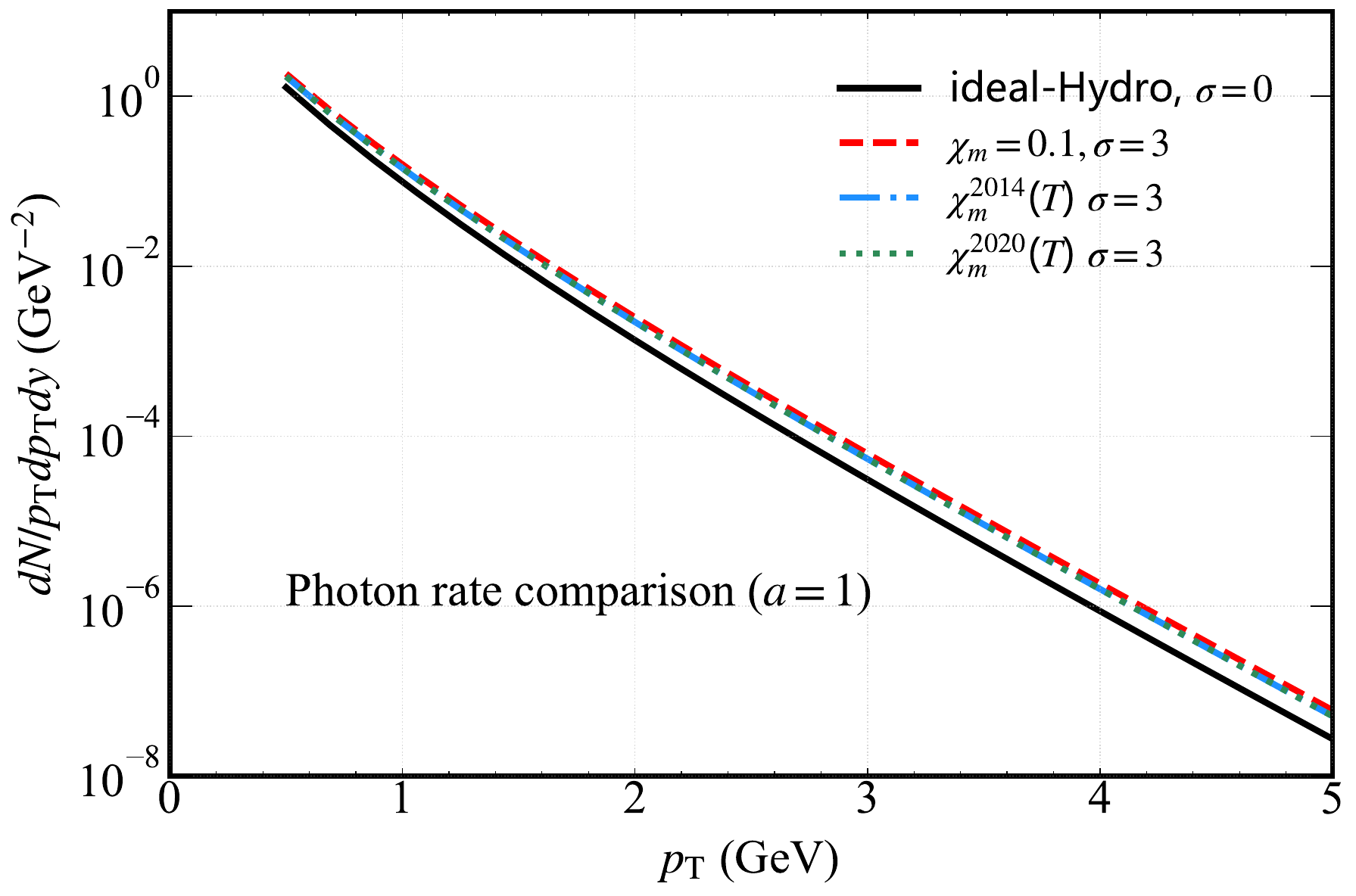}\\
\includegraphics[width=0.85\linewidth]{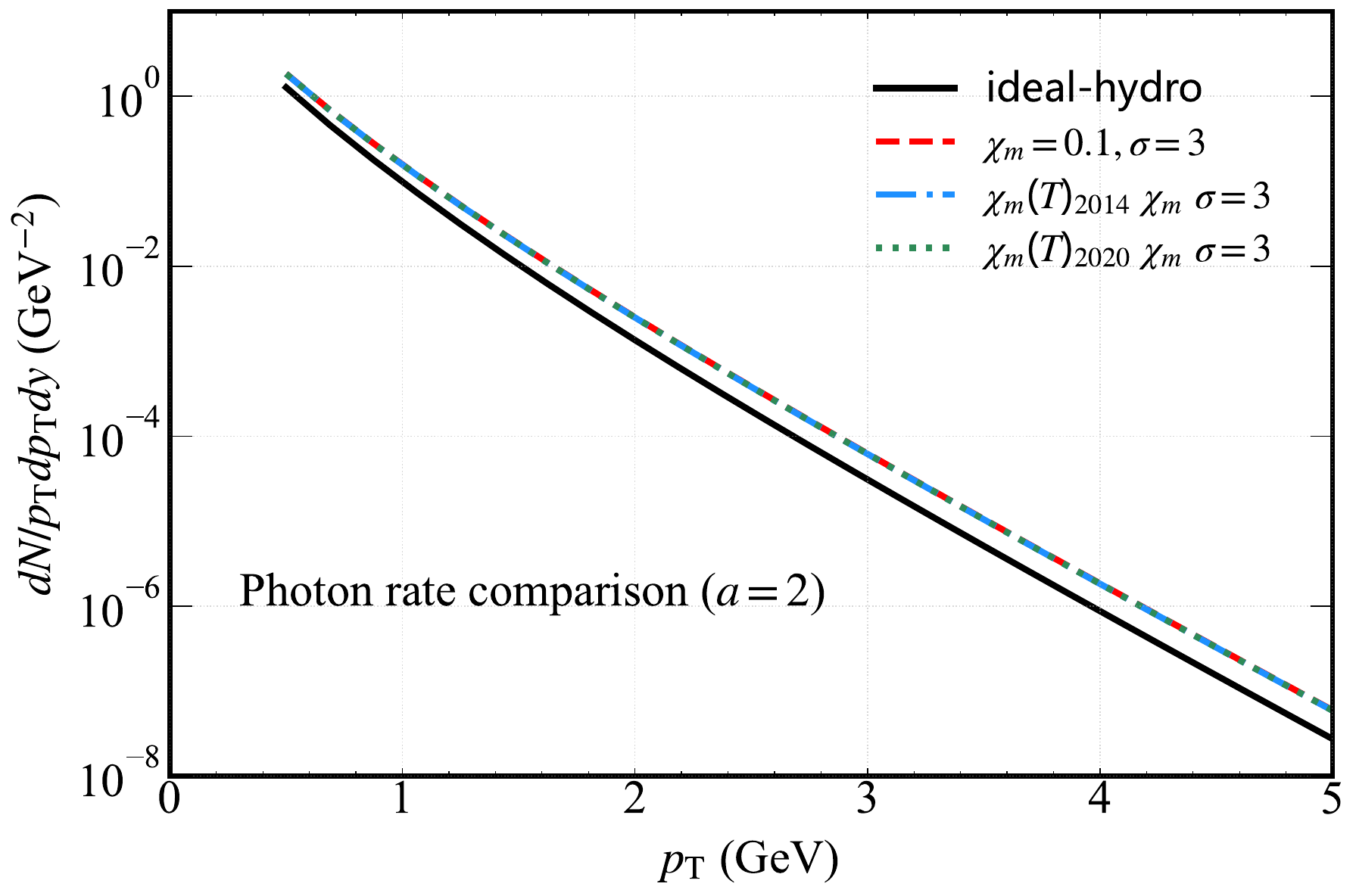}
\caption{(Color online) Total thermal photon production spectra vs. transverse momentum \(p_T\). Upper panel: Comparison of ideal hydrodynamics (\(\sigma=0\)) and magnetized cases with \(\chi_m=0.1\), \(\chi_m=3\), $\chi_{m}^{2014}$($T$), $\chi_{m}^{2020}$($T$) (fixed \(a=1\), \(\sigma=3\)). Lower panel: Same \(\chi_m\) forms and \(\sigma=3\) with fast decay (\(a=2\)).}
\label{f:fig11_chi}
\end{figure}

In the upper panel of Fig.~\ref{f:fig11_chi}, we illustrate how the functional form of QGP magnetic susceptibility---constant (\(\chi_m=0.1\)) versus lattice-QCD-derived temperature-dependent $\chi_{m}(T)$---shapes the thermal photon spectra, with fixed magnetic field decay parameter \(a=1\) (moderate decay) and initial strength \(\sigma=3\). A stark contrast emerges: the field-free ideal hydrodynamics case (\(\sigma=0\)) delivers the lowest yield across all \(p_T\), while all magnetized cases (regardless of \(\chi_m\) form) exhibit distinct enhancement, confirming that even moderate \(\sigma=3\) injects sufficient energy into the QGP to boost photon production. Critically, the spectra for different \(\chi_m\) forms overlap nearly perfectly: no significant differences are observed between constant \(\chi_m=0.1\) and temperature-dependent $\chi_{m}(T)$. This behavior stems from the balanced energy transfer efficiency under \(a=1\): despite differences in how \(\chi_m\) couples the magnetic field to the QGP, the moderate decay rate ensures sufficient interaction time to homogenize the energy input, leading to nearly identical QGP cooling profiles and photon emission windows.

The lower panel of Fig.~\ref{f:fig11_chi} extends this investigation to fast magnetic field decay (\(a=2\)), retaining the same \(\sigma=3\) and set of \(\chi_m\) forms. The trend from the upper panel persists: the ideal hydro case remains the least productive, while all magnetized cases show consistent yield enhancement. More notably, the near-perfect overlap of spectra across different \(\chi_m\) forms is preserved even under fast field decay. This outcome reflects the dominant role of \(a\) in limiting energy transfer: for \(a=2\), the magnetic field decays rapidly, restricting the time window for \(\chi_m\)-mediated coupling. As a result, subtle differences in \(\chi_m\) form are overshadowed by the short interaction duration, leading to uniform QGP thermal evolution and photon yields. Comparing the two panels further reveals that moderate decay (\(a=1\)) yields slightly higher photon rates than fast decay (\(a=2\)) for all magnetized cases---consistent with our prior finding that slower field decay prolongs energy exchange and boosts production.

\begin{figure}[tbp!]
\includegraphics[width=0.85\linewidth]{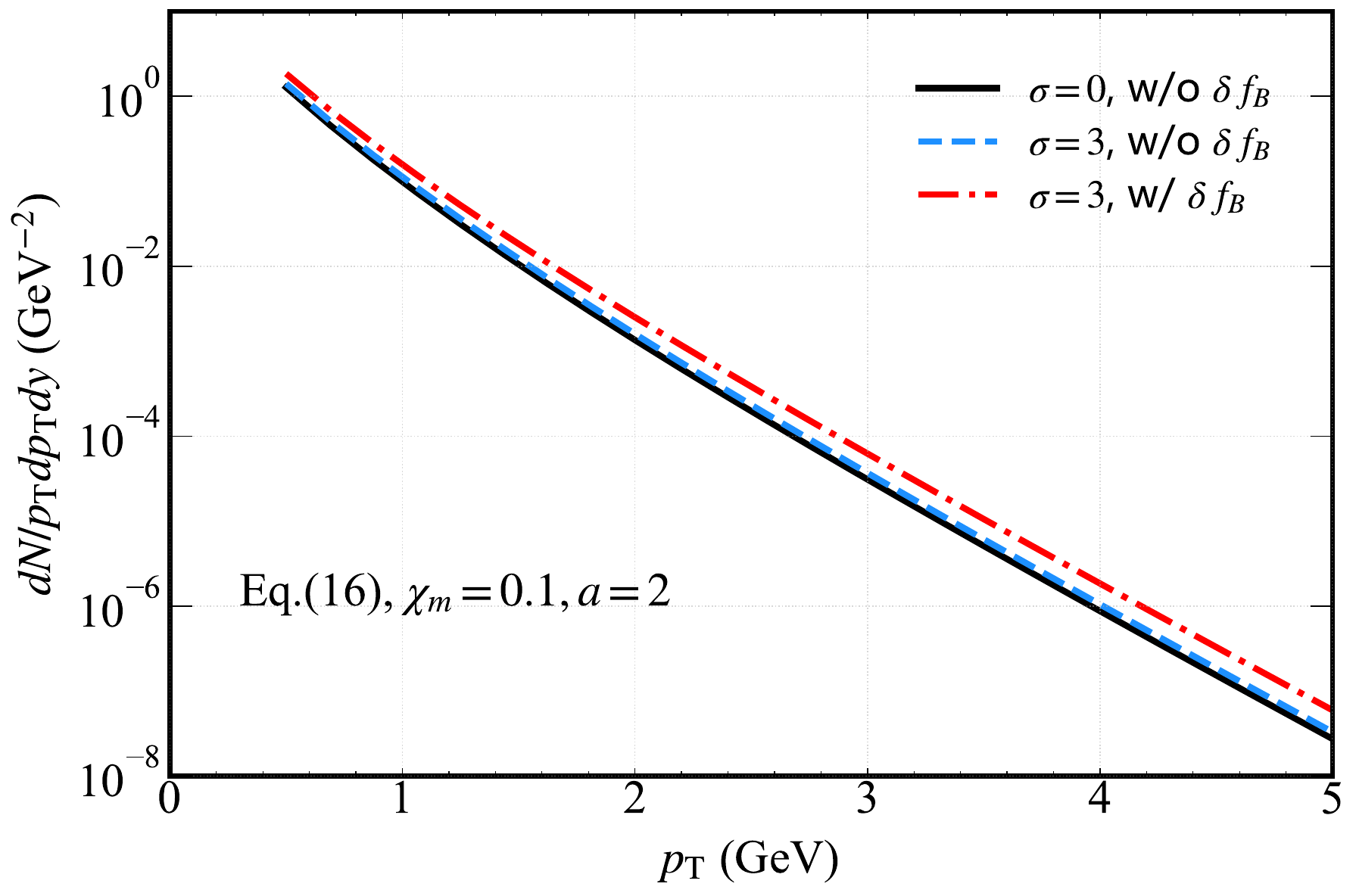}\\
\includegraphics[width=0.85\linewidth]{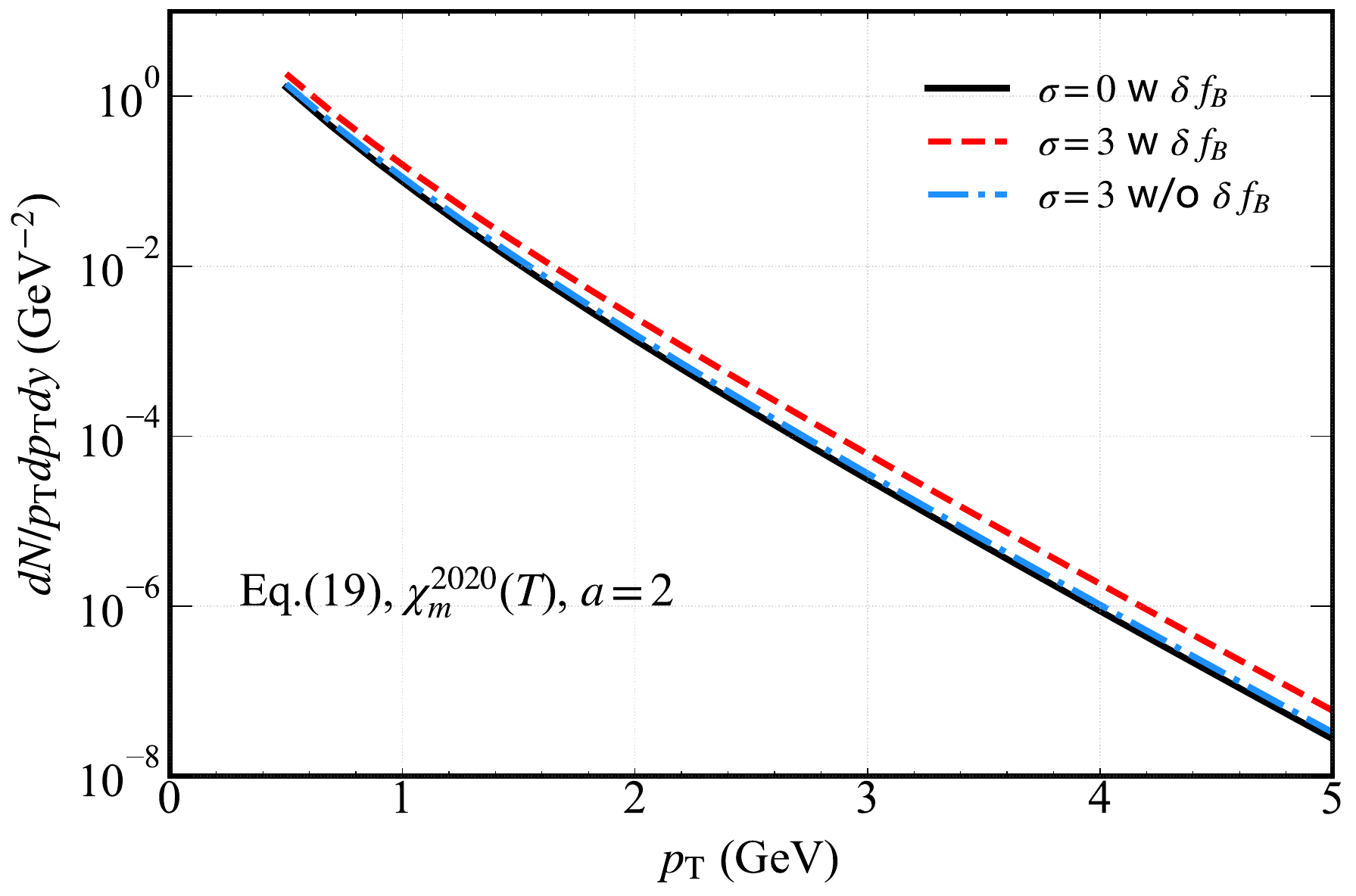}\\
\caption{(Color online)  Total thermal photon production spectra vs. transverse momentum \(p_T\). Upper panel: Comparison of field-free (\(\sigma=0\), without \(f_{\rm EM}\)) and magnetized (\(\sigma=3\), \({\rm with}~f_{\rm EM}\) vs. \({\rm without}~f_{\rm EM}\)) cases with \(\chi_m=0.1\) and \(a=2\). Lower panel: Same \(\sigma\) and \(a=2\) with \(\chi_m^{\rm 2020}(T)\), highlighting the negligible impact of \(f_{\rm EM}\).}
\label{f:fig12_chi}
\end{figure}

In the upper panel of Fig.~\ref{f:fig12_chi}, we highlight a key insight: the magnetic field-induced quark distribution correction (\(f_{\rm EM}\)) enhances thermal photon yields, with fixed \(\chi_m=0.1\), \(a=2\) (fast decay), and \(\sigma=0\) (field-free)/\(\sigma=3\) (magnetized). The \(\sigma=0\) (without \(f_{\rm EM}\)) baseline shows the lowest yield, while the magnetized \(\sigma=3\) case reveals a clear distinction: the spectrum with \( f_{\rm EM}\) corrections delivers measurable enhancement, most pronounced at intermediate \(p_T\) (1-3 GeV). This stems from \( f_{\rm EM}\) modulating quark distributions via field-induced parton momentum shifts, slightly boosting photon-producing collisions (C+A, Bremsstrahlung, A+S) even under weak \(\sigma\) and fast decay.

The lower panel of Fig.~\ref{f:fig12_chi} extends this to the lattice-derived $\chi_{m}^{2020}$($T$) (same \(a=2\), \(\sigma=0/3\)), reinforcing \(f_{\rm EM}\)'s universal yield-enhancing effect. The \(\sigma=0\) baseline remains the least productive; for \(\sigma=3\), both spectra outperform the baseline (confirming \(\sigma\)-driven energy input), and the spectrum with \( f_{\rm EM}\) consistently exceeds that without \( f_{\rm EM}\)---matching the upper panel's trend. This consistency across \(\chi_m\) forms (constant vs. $T$-dependent) underscores \( f_{\rm EM}\)'s enhancement as a robust field-quark interaction effect. While moderate (due to weak \(\sigma\) and fast \(a\)), the non-zero yield margin confirms \( f_{\rm EM}\) is a meaningful correction, refining theoretical predictions by capturing subtle field-modulated parton dynamics.

The results presented thus far employ $\sigma$ values up to 30 to illustrate parametric trends. As discussed in Sec.~\ref{sec:2-A}, the physically expected $\sigma$ at the hydrodynamic initial time $\tau_0$ lies in the range $\sigma \sim 10^{-4}$--$10^{-1}$ (corresponding to $|eB_0|/m_\pi^2 \sim 0.01$--$0.5$), depending on the electrical conductivity of the pre-equilibrium phase. We now discuss the implications of our findings in this realistic parameter window.

\vspace{4pt}
\noindent\textbf{MHD temperature modification.} As shown in Figs.~\ref{f:fig8_chi}--\ref{f:fig11_chi}, the MHD-induced temperature modification is proportional to the factor $(1-a-\chi_m)\sigma$. For realistic $\sigma \lesssim 0.1$, this term is suppressed by at least an order of magnitude relative to the benchmark $\sigma=1$ shown in the upper panel of Fig.~\ref{f:fig8_chi}, where the difference from the $\sigma=0$ case is already marginal. We therefore conclude that the MHD temperature modification through $\chi_m$ is indeed negligible for realistic field strengths. We explicitly acknowledge that the $\chi_m$-dependent MHD temperature effect alone does not produce observable modifications to the photon spectrum at realistic $\sigma$.

\vspace{4pt}
\noindent\textbf{The $f_{\rm EM}$ correction.} The situation is different for the $f_{\rm EM}$ correction. Since $f_{\rm EM}\propto B \propto \sqrt{\sigma}$, the EM-induced photon yield $dN_{\rm EM}/d^2p_T dy$ scales as $\sqrt{\sigma}$ at leading order. For $\sigma=0.1$ ($|eB_0|/m_\pi^2 \approx 0.5$), the $f_{\rm EM}$ contribution is reduced by a factor $\sqrt{0.1/3} \approx 0.18$ compared to the $\sigma=3$ case shown in Fig.~\ref{f:fig12_chi}. This is still non-negligible and within the reach of high-precision measurements. Moreover, as demonstrated in Refs.~\cite{Sun:2023pil,Sun:2023rhh}, the key observable impact of $f_{\rm EM}$ lies not in the photon yield enhancement but in the azimuthal anisotropy it induces. The photon elliptic flow induced by $f_{\rm EM}$ is $v_2^{\rm EM} \approx 0.5$--$0.6$, largely independent of the field magnitude, because it arises from the geometric coupling between the magnetic field direction and the rapidity-odd dipole moment of the medium. Even for $\sigma \sim 10^{-2}$, the $f_{\rm EM}$ correction to $v_2^\gamma$ can be significant if the background photon $v_2$ is small. Quantifying this effect within the present MHD framework requires event-by-event (3+1)-dimensional hydrodynamic simulations~\cite{Sun:2023pil}, which we defer to a dedicated future study.

\begin{figure}[tbp!]
\includegraphics[width=0.85\linewidth]{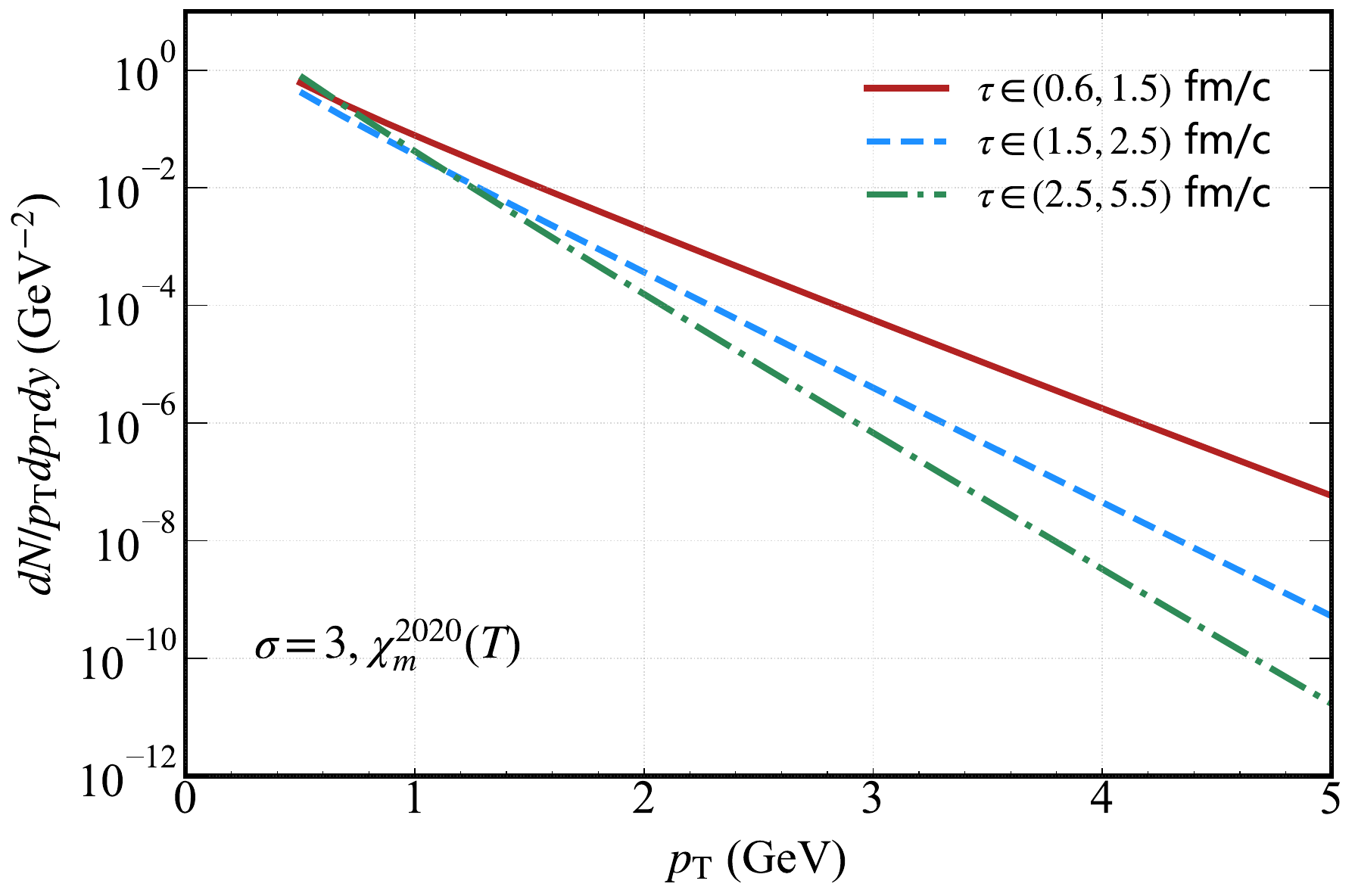}
\caption{(Color online) Thermal photon production from the QGP in an external magnetic field includes the total yield and contributions from distinct proper time ($\tau$) intervals.}
\label{f:fig13_chi}
\end{figure}

In Fig.~\ref{f:fig13_chi}, we present thermal photon spectra for three proper time intervals [(0.6, 1.5) fm/c, (1.5, 2.5) fm/c, (2.5, 5.5) fm/c] under \(\sigma=3\). We find that low \(p_T\) photons (\(p_T \leq 1.5\) GeV) draw non-negligible contributions from all QGP stages, as their production relies on low-energy soft processes (e.g., soft Compton scattering, low-threshold bremsstrahlung). In contrast, high \(p_T\) photons (\(p_T \geq 2\) GeV) are overwhelmingly dominated by the earliest, ultra-hot interval [(0.6, 1.5) fm/c], where high-momentum-transfer hard processes (e.g., hard parton collisions, high-energy \(q\bar{q}\) annihilation) thrive due to maximized parton collision rates and thermal energy, while cooling in middle and late stages sharply suppresses such high \(p_T\) production. This distinction renders thermal photons a dual-purpose probe: low \(p_T\) encodes the full QGP lifetime, while high \(p_T\) offers a direct window into its early, high-temperature phase.

\section{Conclusions}  
\label{section-5}
In this work, we have extended the investigation of thermal photon production in the quark-gluon plasma (QGP) to the magnetohydrodynamic (MHD) regime, incorporating magnetic susceptibility (\(\chi_m\)) and the magnetic field-induced quark distribution correction (\(f_{\rm EM}\)). This extension deepens our understanding of how magnetized QGP dynamics modulates electromagnetic signatures.

We adopt the well-known relativistic MHD framework~\cite{Pu:2016ayh} to describe the QGP energy-momentum tensor under external magnetic fields, incorporating \(\chi_m\) in both constant and lattice-derived temperature-dependent forms ($\chi_{m}^{2014}$($T$) and $\chi_{m}^{2020}$($T$)) to capture the medium's response to magnetic fields. Building upon the Pu-Bjorken flow assumption~\cite{Roy:2015kma,Pu:2016ayh,Jiang:2024mts,Peng:2022cya}, we derive analytical solutions for the QGP temperature evolution that explicitly account for \(\chi_m\), thereby revealing how fluid-field coupling modulates thermal cooling.

Another key extension of this work is the inclusion of \(f_{\rm EM}\)-the magnetic field-induced correction to quark distribution functions-which captures field-modulated parton momentum shifts and occupation numbers. We calculate thermal photon production rates for three dominant processes~\cite{Traxler:1995kx,Steffen:2001pv,Bhatt:2010cy}: Compton scattering with \(q\bar{q}\) annihilation (C+A), bremsstrahlung (Bre), and \(q\bar{q}\) annihilation with additional scattering (A+S). The total photon spectrum is obtained by integrating these rates over the QGP spacetime evolution, incorporating both \(\chi_m\)-dependent temperature profiles and \(f_{\rm EM}\) corrections.

Our results uncover several critical insights regarding magnetized QGP photon production: (i) The initial magnetic field strength (\(\sigma\)) and decay exponent (\(a\)) remain the dominant regulators of photon yield via MHD temperature modification, wherein stronger \(\sigma\) and slower decay (smaller \(a\)) enhance yields by prolonging energy exchange between the field and the QGP, thereby slowing thermal cooling and extending emission windows. At realistic $\sigma$ values ($\sigma \lesssim 0.1$, corresponding to $|eB_0|/m_\pi^2 \lesssim 0.5$ at $\tau_0$), this MHD temperature effect becomes negligible, as we have explicitly verified. (ii) Magnetic susceptibility \(\chi_m\), whether in constant or realistic temperature-dependent form, exerts negligible influence on photon yields under the studied conditions, as its modulation of fluid-field coupling is overshadowed by the effects of \(a\) and \(\sigma\). This finding, which aligns with the expectation that $\chi_m \sim 0.01$--$0.1$ is too small to affect the bulk temperature evolution appreciably, is in fact one of the quantitative outcomes of our systematic survey. (iii) The \(f_{\rm EM}\) correction, though moderate in magnitude, delivers a measurable yield enhancement (especially pronounced at intermediate \(p_T\)), validating its role as a meaningful refinement for precise theoretical predictions. Importantly, the $f_{\rm EM}$ correction is linear in the magnetic field $B$, not in the MHD-modified temperature, and it generates substantial photon anisotropy that is largely independent of the field magnitude. This makes it a potentially observable signature even at realistic $\sigma$. (iv) Low-\(p_T\) photons (\(p_T \leq 1.5\) GeV) receive contributions from all stages of QGP evolution, whereas high-\(p_T\) photons (\(p_T \geq 2\) GeV) are dominated by the early, ultra-hot phase, reflecting the distinct energy thresholds of soft and hard photon-producing processes.

We emphasize that the primary contribution of this work is methodological: it establishes, for the first time, a unified computational framework that simultaneously incorporates MHD temperature evolution with magnetic susceptibility $\chi_m$, the weak-field quantum correction $f_{\rm EM}$ to quark distributions, and the complete set of photon production channels (C+A, Bremsstrahlung, A+S) within a single, reproducible calculation. The systematic parameter scan over $\sigma$, $a$, and $\chi_m$ serves to disentangle the relative importance of these effects and to provide a benchmark for future, more sophisticated studies. While the $\chi_m$-driven MHD temperature modification is shown to be negligible for realistic field strengths---confirming the physical intuition that a weak magnetic field does not appreciably alter the bulk medium evolution---the unified framework itself, and in particular the $f_{\rm EM}$ implementation, lays the necessary groundwork for extending these calculations to (3+1)-dimensional dissipative MHD, where the interplay between magnetic fields, longitudinal dynamics, and photon anisotropy can be fully quantified.

This study advances prior work by systematically integrating magnetic susceptibility \(\chi_m\) and weak magnetic field \(f_{\rm EM}\) corrections into the MHD framework for QGP photon production, providing a more comprehensive description of field-parton-thermal dynamics. The simplicity and reproducibility of this approach lay the groundwork for further explorations, including viscous magnetohydrodynamics~\cite{Jiang:2024mts}, longitudinally accelerated magnetohydrodynamics~\cite{She:2019wdt}, spin-hydrodynamics~\cite{Hattori:2019lfp,Fukushima:2020ucl,Hongo:2021ona,Li:2020eon,She:2021lhe,Daher:2022xon,Biswas:2023qsw,Peng:2021ago,Florkowski:2017ruc,Florkowski:2018fap,Li:2019qkf,Bhadury:2020puc,Shi:2020htn,Hongo:2022izs,Weickgenannt:2022zxs,Bhadury:2022ulr,Weickgenannt:2022qvh,Gallegos:2021bzp,Gallegos:2022jow,Montenegro:2020paq,She:2024rnx,She:2025qri,Wang:2021wqq}, spin-magnetohydrodynamics~\cite{Fang:2024sym,Bhadury:2022ulr}, non-extensive magnetohydrodynamics~\cite{Shen:2017pyo}, as well as lepton physics~\cite{Wu:2024vyc}-all of which could reveal richer interactions between magnetic fields, parton dynamics, and thermal photon emission. These investigations will not only deepen our understanding of QGP electromagnetic signatures but also provide theoretical predictions for heavy-ion collision experiments aiming to probe the transport and magnetic response properties of the QGP.

\begin{acknowledgements}
This work was supported by the National Natural Science Foundation of China (NSFC) under Grant Nos.~12305138, the Natural Science Foundation of Hubei Province No.~2026AFB678. Duan She's research is funded by the Startup Research Fund of Henan Academy of Sciences (No. 231820058), the 2024 Henan Province International Science and Technology Cooperation Projects (No. 242102521068), the High-level Achievements Reward and Cultivation Projects (No. 20252320001),
and the Key Laboratory of Quark and Lepton Physics Contracts No. QLPL2025P01.
\end{acknowledgements}

\bibliographystyle{unsrt}
\bibliography{main}

\appendix
\section{Perturbative Solution of MHD}
\label{app:perturbative_mhd}
We derive the approximate solution for the MHD temperature evolution equation in the case of the lattice-derived magnetic susceptibility $\chi_{m}^{2014}$($T$), starting from the governing equation:
\begin{equation}
\begin{aligned}
\frac{\partial T}{\partial\tau} &+\frac{(\kappa+1)T}{4\kappa\tau} \\
&+\left[1-a-\frac{e^2}{3\pi^2}\log\left(\frac{T}{0.11}\right)\right]\frac{\sigma T_0^4\tau_0^{2a}}{4\kappa a_1 T^3\tau^{2a+1}} = 0. 
\end{aligned}
\label{eq:mhd_governing}
\end{equation}
Introducing the dimensionless normalized temperature \(\widetilde{T} = T/T_0\) (with initial condition \(T(\tau_0)=T_0\)), Eq.~\eqref{eq:mhd_governing} can be rewritten as:
\begin{equation}
\begin{aligned}
\frac{\partial\widetilde{T}}{\partial\tau}&+\frac{(\kappa+1)\widetilde{T}}{4\kappa\tau} \\ 
&+\left[1-a-\frac{e^2}{3\pi^2}\log\left(\frac{\widetilde{T}T_0}{0.11}\right)\right]\frac{\sigma\tau_0^{2a}}{4\kappa a_1 \widetilde{T}^3\tau^{2a+1}} = 0.
\label{eq:mhd_normalized}
\end{aligned}
\end{equation}

We employ a nonconserved charge method, solving for \(\widetilde{T}(\tau)\) via the auxiliary equation:
\begin{equation}
\frac{d}{d\tau}f(\tau) + m\frac{f(\tau)}{\tau} = f(\tau)\frac{d}{d\tau}\lambda(\tau),
\label{eq:auxiliary_eq}
\end{equation}
where \(m\) is a constant and \(\lambda(\tau)\) is a known function. The general solution of Eq.~\eqref{eq:auxiliary_eq} is:
\begin{equation}
f(\tau) = f(\tau_0)\exp\left[\lambda(\tau)-\lambda(\tau_0)\right]\left(\frac{\tau_0}{\tau}\right)^m,
\label{eq:auxiliary_solution}
\end{equation}
with \(f(\tau_0)\) determined by the initial condition at proper time \(\tau_0\).

Matching Eq.~\eqref{eq:mhd_normalized} to Eq.~\eqref{eq:auxiliary_eq}, we identify \(m=(\kappa+1)/(4\kappa)\) and:
\begin{equation}
\frac{d\lambda}{d\tau} = -\left[1-a-\frac{e^2}{3\pi^2}\log\left(\frac{\widetilde{T}T_0}{0.11}\right)\right]\frac{\sigma\tau_0^{2a}}{4\kappa a_1 \widetilde{T}^4\tau^{2a+1}}.
\label{eq:lambda_deriv}
\end{equation}
Using Eq.~\eqref{eq:auxiliary_solution} with \(\widetilde{T}(\tau_0)=1\), the formal solution for \(\widetilde{T}(\tau)\) is:
\begin{equation}
\widetilde{T}(\tau) = \left(\frac{\tau_0}{\tau}\right)^{\frac{\kappa+1}{4\kappa}}x(\tau),
\label{eq:tildeT_x}
\end{equation}
where \(x(\tau) = \exp\left[\lambda(\tau)-\lambda(\tau_0)\right]\) satisfies \(x(\tau_0)=1\) and \(dx = x d\lambda\).

Substituting Eq.~\eqref{eq:tildeT_x} into Eq.~\eqref{eq:lambda_deriv} yields the evolution equation for \(x(\tau)\):
\begin{equation}
\begin{aligned}
\frac{dx}{d\tau} = &-x\left[1-a-\frac{e^2}{3\pi^2}\log\left(\frac{T_0}{0.11}\left(\frac{\tau_0}{\tau}\right)^{\frac{\kappa+1}{4\kappa}}x\right)\right] \\
&\times\frac{\sigma\tau_0^{2a}}{4\kappa a_1 \left(\frac{\tau_0}{\tau}\right)^{\frac{\kappa+1}{\kappa}}x^3\tau^{2a+1}}.
\label{eq:dx_dtau}
\end{aligned}
\end{equation}

Treat \(\epsilon = \sigma/a_1\) as a small parameter, expanding \(x(\tau)\) to \(\mathcal{O}(\epsilon)\):

- \textit{Zero-order} (\(\mathcal{O}(\epsilon^0)\)): Eq.~\eqref{eq:dx_dtau} simplifies to \(dx/d\tau=0\), giving \(x_0(\tau)=1\) (via \(x(\tau_0)=1\)).

- \textit{First-order} (\(\mathcal{O}(\epsilon^1)\)): Substitute \(x(\tau)\approx x_0=1\) into Eq.~\eqref{eq:dx_dtau} and integrate from \(\tau_0\) to \(\tau\):
\begin{equation}
\begin{aligned}
    x_1(\tau) = 1 + &
    \int_{\tau_0}^{\tau}\left[-1+a+\frac{e^2}{3\pi^2}\log\left(\frac{T_0}{0.11}\left(\frac{\tau_0}{\tau_1}\right)^{\frac{\kappa+1}{4\kappa}}\right)\right] \\
    &\times \frac{\epsilon\tau_0^{2a}}{4\kappa \left(\frac{\tau_0}{\tau_1}\right)^{\frac{\kappa+1}{\kappa}}\tau_1^{2a+1}}d\tau_1.
\end{aligned}
\label{eq:x1_integral}
\end{equation}

\begin{figure}[tbp!]
\includegraphics[width=0.9\linewidth]{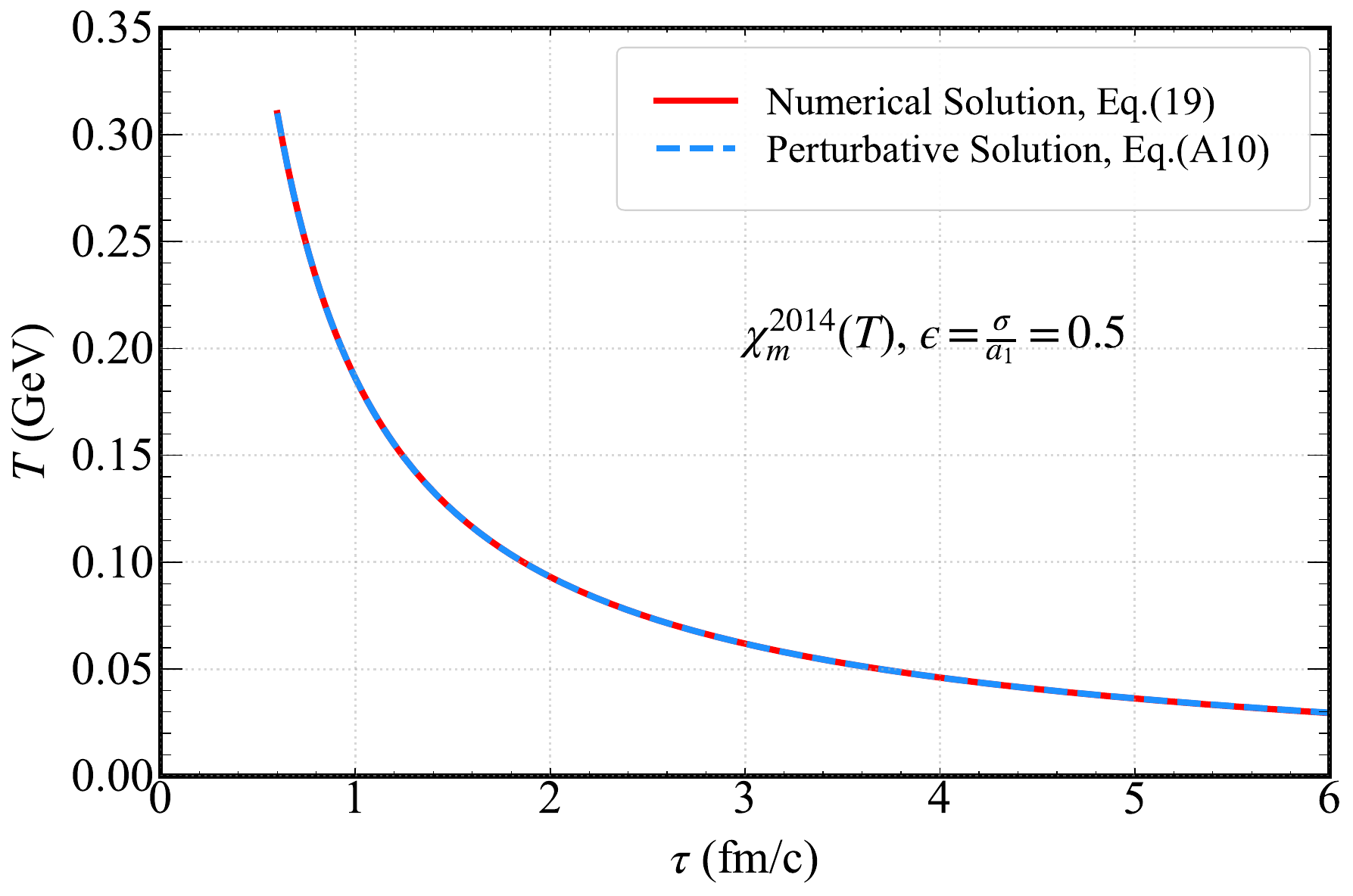} 
\caption{(Color online) Temperature dependence of the magnetic susceptibility $\chi_{m}(T)$ for three prescriptions: constant $\chi_{m}=0.01$, the 2014 lattice QCD result (Eq.~(\ref{eq:chi_2014})), and the 2020 lattice QCD parametrization (Eq.~(\ref{eq:chi_2020})), over 100 $\leq~T~\leq$ 350 MeV.}
\label{f:chim_c_1}
\end{figure}

Using the substitution \(\tau_2 = \tau_1/\tau_0\) (\(d\tau_1 = \tau_0 d\tau_2\)) and splitting the integral into convergent limits, we compute:
\begin{equation}
\begin{aligned}
x_1(\tau) &= 1 + \frac{\epsilon}{48\pi^2(2a\kappa-\kappa-1)^2}\Bigg\{-e^2(1+\kappa) \\
&\quad+ 12(a-1)\pi^2(2a\kappa-\kappa-1)\\
&\quad + 4e^2(2a\kappa-\kappa-1)\log\left(\frac{T_0}{0.11}\right) - \left(\frac{\tau_0}{\tau}\right)^{2a-\frac{\kappa+1}{\kappa}}\Bigg[\\
&\quad -e^2(\kappa+1) + 12(a-1)\pi^2(2a\kappa-\kappa-1)\\
&\quad + 4e^2(2a\kappa-\kappa-1)\log\left(\frac{T_0}{0.11}\left(\frac{\tau_0}{\tau}\right)^{\frac{1+\kappa}{4\kappa}}\right)\Bigg]\Bigg\},
\end{aligned}
\label{eq:x1_solution}
\end{equation}
valid for \(2a\kappa-\kappa-1\neq0\).

Substituting \(x_1(\tau)\) back into Eq.~\eqref{eq:tildeT_x}, the perturbative solution for \(T(\tau)\) is:
\begin{equation}
T(\tau) = T_0\left(\frac{\tau_0}{\tau}\right)^{\frac{\kappa+1}{4\kappa}}x_1(\tau),
\label{eq:final_T_solution}
\end{equation}
where \(x_1(\tau)\) is given by Eq.~\eqref{eq:x1_solution}. This solution is very stable for $a=1$ and small \(\epsilon=\sigma/a_1\).

Fig. ~\ref{f:chim_c_1} summarizes the comparison of QGP temperature evolution \(T(\tau)\) (in GeV) versus proper time \(\tau\) (in fm/c) between the perturbative solution (Eq.(\ref{eq:final_T_solution})) and numerical solution (Eq. (\ref{eq:T-1})), under the condition $\tau_{0}=0.6$ fm, $T_{0}=0.31$ GeV, $a=1$, $\kappa=3$ and \(\epsilon = \sigma/a_1 = 0.5\). 

Within the studied \(\tau\) range (0.6-6 fm/c), the two solutions show very good consistency: both exhibit the physically expected monotonic decay of QGP temperature with increasing proper time, a signature of QGP expansion and cooling in relativistic heavy-ion collisions. This agreement validates the key assumption-treating \(\epsilon\) as a small parameter for perturbative expansion-confirming that the perturbative solution is reliable when \(\epsilon\) takes small values. 

Notably, while the main text prioritizes the numerical solution for its general applicability across different \(\epsilon\) regimes, the consistent results here demonstrate that the perturbative solution (Eq.(~\ref{eq:final_T_solution})) can serve as a concise and accurate alternative for cases with small \(\epsilon\). It avoids the computational complexity of numerical simulations while preserving physical accuracy, and serving as an efficient supplementary tool for analyses of relevant thermodynamic properties and photon production rates.

\end{document}